\documentclass[smallextended]{svjour3}       
\smartqed  
\usepackage{graphicx}
\usepackage{epsfig}
\usepackage{pdfpages}
\usepackage{subfigure}
\usepackage{marvosym} 
\usepackage{color}
\usepackage{ulem}

\usepackage{amsfonts}
\usepackage{amsmath}
\usepackage{amssymb}

\usepackage[title]{appendix}

\usepackage{sidecap}

\begin{document}

\title{Pareto-optimal solution for the quantum battle of the sexes}

\author{A. Consuelo-Leal$^{1}$\Letter  \and	
        A. G. Araujo-Ferreira$^{1} $ 	\and	
        E. Lucas-Oliveira$^{1} $	 	\and	
        T. J. Bonagamba$^{1} $		 	\and	
        R. Auccaise$^{2} $			 			
}

\institute{Corresponding author :  A. Consuelo-Leal \\ \email{adrianeleal@ifsc.usp.br} \\ 
           $^{1} $ Instituto de F\'{i}sica de S\~{a}o Carlos, Universidade de S\~{a}o Paulo, CP 369, 13560-970, S\~{a}o Carlos, SP, Brasil. \\        
           $^{2} $ Departamento de F\'{i}sica, Universidade Estadual de Ponta Grossa, Av. Carlos Cavalcanti 4748, 84030-900 Ponta Grossa, PR, Brasil.
}


\date{Received: date / Accepted: date}

\maketitle

\begin{abstract}
Quantum games have gained much popularity in the last two decades. Many of these quantum games are a redefinition of iconic classical games to fit the quantum world, and they gain many different properties and solutions in this different view. In this letter, we attempt to find a solution to an asymmetric quantum game which still troubles quantum game researchers, the quantum battle of the sexes. To achieve that, we perform an analysis using the Eisert-Wilkens-Lewenstein protocol for this asymmetric game. The protocol highlights two solutions for the game, which solve the dilemma and satisfy the Pareto-optimal definition, unlike previous reports that rely on Nash equilibrium. We perform an experimental implementation using the NMR technique in a two-qubit system. Our results eliminate dilemmas on the quantum battle of the sexes and provide us with arguments to elucidate that the Eisert-Wilkens-Lewenstein protocol is not restricted to symmetric games when at the quantum regime.
\keywords{Quantum game \and quantum strategy \and symmetric and asymmetric game \and pareto-optimal \and quantum circuit \and Nuclear magnetic resonance}
\end{abstract}

\section{Introduction}

Human decision making is a process in which an alternative is selected rationally or intuitively. Rational decisions can be studied formally and in game theory are part of a player's strategy. The strategy determines every action {that players} may take with the goal of maximizing his chance to win or maximize {their} payoffs. From a collection of a few seemingly simple set of rules of payoff and possible strategy choices, many emblematic games were developed, analyzed and studied such as the prisoner's dilemma, the battle of the sexes, coin tossing, rock-scissors-paper  \cite{smith1982Book,hofbauer1998Book} and others  \cite{conway2001Book}.

At the end of the 90's, the classical interpretation of those games started their extension to the quantum world. In that sense, the Penny PQ flip over \cite{meyer1999}, gambling \cite{goldenberg1999} and prisoner's dilemma \cite{eisert1999} were the first games analyzed in the quantum regime. From those cited, the most versatile problem corresponds to the prisoner's dilemma, a symmetrical game successfully studied by the protocol proposed by Eisert-Wilkens-Lewenstein \cite{eisert1999}, which had its discussion extended to multiple players \cite{johnson2001,benjamin2001PRA}, concepts of non-locality \cite{brunner2013,pappa2015,melo-Luna2017}, entanglement \cite{lindsell2014}, concepts of symmetric and asymmetric games \cite{situ2015}, or in the context of Bayesian games \cite{solmeyer2017}   and was verified experimentally by NMR techniques using a cytosine sample \cite{du2002} and by optical quantum circuits on two and four-qubit systems  \cite{melo-Luna2017,prevedel2007,balthazar2015JPB} and using ion-trap setup on five-qubit system \cite{solmeyer2018}.

Another critical contribution to quantum games is the discussion started by Marinatto and Weber focused on the quantum of the battle of the sexes \cite{marinatto2000}. Although Marinatto-Weber's discussion brought out an interesting point of view, their analysis restricts the choices of the players as pointed out by Benjamin \cite{benjamin2000}. Numerous theoretical discussions attempted to show its potentiality \cite{du2001ppMarco,du2001ppNovembro,fariasneto2004pp,alonso-sanz2014,alonso-sanz2015}, highlighting the use of a general initial quantum state \cite{nawaz2004}, arguing the omission of a disentangling quantum  gate   $\widehat{\mathbf{J}}^{\dag }$ to maintain the quantum game in its highest correlated regime where the dilemma does not exist \cite{benjamin2000,iqbal2002}, using the Harsanyi-Selten algorithm to accomplish an ultimate solution \cite{frackiewicz2009} and playing asymmetric coordination games \cite{situ2014}. Out of these previously cited, the most important is about the omission of a disentangling quantum gate.  Specifically, we {draw attention} to the last two-qubit gate of the Eisert-Wilkens-Lewenstein protocol, the disentanglement operator  $\widehat{\mathbf{J}}^{\dag }$,    because it highlights the distinction between the Marinatto-Weber  and the Eisert-Wilkens-Lewenstein protocols. This argument is underlined by Benjamin \cite{benjamin2000} who points out that the second inverse gate  $\widehat{\mathbf{J}}^{\dag }$ ensures that the classical game remains embedded within the quantum game  \cite{benjamin2001PRA}, while Melo-Luna claims that it belongs to the measurement procedure \cite{melo-Luna2017}.

From the above arguments, in this paper, we present a theoretical analysis and experimental verification of an asymmetrical game considering the application of the Eisert-Wilkens-Lewenstein protocol \cite{eisert1999}. In order to achieve this task, we organize this paper as follows: in section \ref{sec:ClassicalGame}  we introduce  a simple, accurate description of the classical version of the battle of the sexes game and in section \ref{sec:QuantumGame} we extend those concepts to its quantum version. In section \ref{sec:Discussion} we analyse the theoretical data, applying  fundamentals of game theory, mathematical definitions and experimental procedures. In section \ref{sec:Conclusions} we present our conclusions. At the end of the manuscript, we show in appendix \ref{sec:ExperimentalDetails} an extended experimental description and  in appendix \ref{sec:TheoreticalAnalysis} an extra mathematical detail.

\section{Classical game}\label{sec:ClassicalGame}

Two players, Alice and Bob, need to decide their entertainment for a Saturday night. Both have two options they can pick, and each one must choose simultaneously without any communication between them. The possibilities are either to go to the Opera  $\left( O \right)$  or to watch Television  $\left( T \right)$. Alice loves the Opera, but Bob would prefer to stay at home and watch television. However, both of them want to stay together rather than going to their favored activity. Thus, the dilemma arises. They want to maximize their happiness with their choices, and the result of their decisions can be visualized on the bi-matrix of payoffs on Tab \ref{tab:PayoffsBattleSexes}. If both of them choose Television (Opera), Alice's payoff is $\beta \ \left( \alpha \right)$, and Bob's payoff is $\alpha \ \left( \beta \right)$. If Alice chooses Opera (Television) and Bob Television (Opera), then both of their payoffs is  $\gamma $.

From the fundamentals of game theory \cite{marinatto2000,dixit2014Book}, and for a two player game (Alice and Bob), a pair of strategies $\left( s_{\text{A}}^{\star },s_{\text{B}}^{\star }\right)$ is defined as \textit{Nash equilibrium}, if for both payoff functions $\overline{\$}_{\text{A}}$ and $\overline{\$}_{\text{B}}$ the pair  $\left( s_{\text{A}}^{\star },s_{\text{B}}^{\star }\right)$  satisfies the following inequalities
\begin{eqnarray}
\overline{\$}_{\text{A}}\left( s_{\text{A}}^{\star },s_{\text{B}}^{\star }\right)  &\geq &\overline{\$}_{\text{A}} \left( s_{\text{A}},s_{\text{B}}^{\star }\right) \text{,} \label{NashEquiAlice} \\
\overline{\$}_{\text{B}}\left( s_{\text{A}}^{\star },s_{\text{B}}^{\star }\right)  &\geq &\overline{\$}_{\text{B}} \left( s_{\text{A}}^{\star },s_{\text{B}}\right) \text{,} \label{NashEquiBob}
\end{eqnarray}
{where   $s_{\text{A,B}}$ means the player's choice from a set of options and   $s_{\text{A,B}}^{\star }$ means the most efficient player's choice obeying the game rules that maximize their payoffs. }
From these equations and the payoffs table, we ascertain that Television-Television and Opera-Opera satisfy the condition of Nash equilibrium   \cite{dixit2014Book}, although the payoffs for each player are different in both cases. This difference is the main characteristic of an asymmetrical game.

\begin{table}[!ht]
\begin{center}
\begin{tabular}{llccc}
\hline\hline
     \multicolumn{2}{c}{ }   &           \multicolumn{3}{c}{Bob}  \\ 
     \multicolumn{2}{c}{ }   &  $T$ & & $O$ \\  \hline
					&	\quad	$T$	\quad	&	\quad	$ \left( \beta , \alpha \right)$	\quad	& &	\quad	$ \left( \gamma, \gamma \right)$	\quad	\quad		\\
\quad Alice \quad	&						&									& &                                  \\
					&	\quad	$O$	\quad	&	\quad	$ \left( \gamma, \gamma\right)$	\quad	& &	\quad	$ \left( \alpha, \beta \right)$	\quad		\quad	\\ [3pt]
\hline \hline
\end{tabular}
\caption{Bi-matrix of payoffs  for the battle of the sexes game at its classical regime \cite{smith1982Book,hofbauer1998Book}. The first (second) entry in the parenthesis denotes Alice's (Bob's) payoffs. Parameters values obey the relation $ \alpha > \beta > \gamma $. }	\label{tab:PayoffsBattleSexes}
\end{center}
\end{table}

\section{Quantum game}\label{sec:QuantumGame}

Now, let us assume that the quantum battle of the sexes obeys the Eisert-Wilkens-Lewenstein's proposal \cite{eisert1999} and follow Benjamin's suggestion \cite{benjamin2000}. Mathematically, the two-player game is defined in a Hilbert space generated from the standard computational basis for a two-qubit system, $\left\{ \left\vert O_{\text{A}},O_{\text{B}}\right\rangle ,\left\vert T_{\text{A}},O_{\text{B}}\right\rangle ,\left\vert O_{\text{A} },T_{\text{B}}\right\rangle ,\left\vert T_{\text{A}},T_{\text{B} }\right\rangle \right\} $, such that for a one qubit state we define the kets:
\[
\left\vert O\right\rangle \equiv \left\vert 0\right\rangle =\left[ 
\begin{array}{c}
1 \\ 
0
\end{array}
\right] \text{,\qquad }\left\vert T\right\rangle \equiv \left\vert
1\right\rangle =\left[ 
\begin{array}{c}
0 \\ 
1
\end{array}
\right] \text{,}
\]
in which $\left\vert O\right\rangle $ encodes the classical \textit{O}pera strategy, and $\left\vert T\right\rangle $ encodes the classical \textit{T}elevision strategy. {Given the quantum mechanical nature of the system, this encoding procedure generates many other possibilities through the principle of superposition of quantum states. The quantum state superposition is achieved performing unitary transformations, and they are mathematically defined by operators. These operators depend on angular parameters which can be mapped onto the concept of strategy in game theory context. In this sense, we define the quantum operator that generates quantum states representing strategies.}
\begin{figure}[!ht]
\begin{center}
\includegraphics[width=0.80\textwidth]{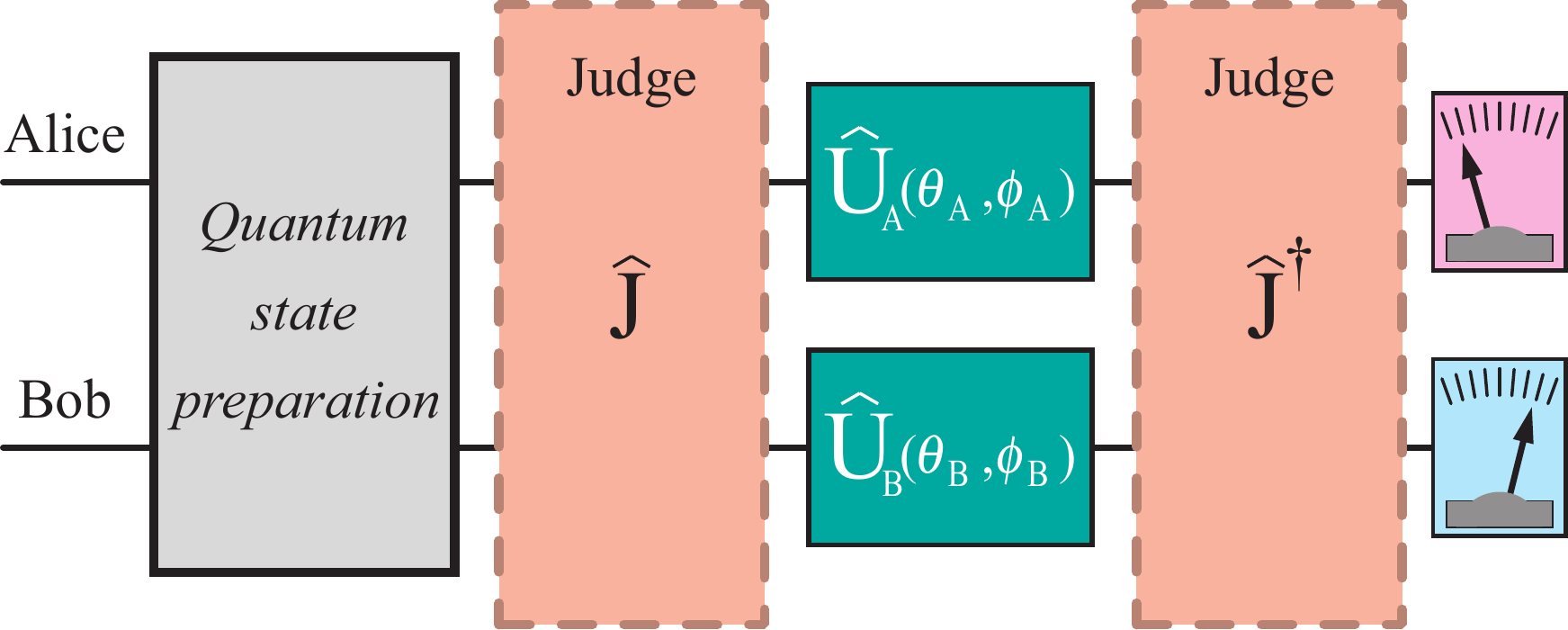}
\end{center}
\caption{(Color online) Two-player quantum circuit describing Eisert-Wilkens-Lewenstein protocol \cite{eisert1999}. The gray box represents the quantum state preparation $ \left\vert 00\right\rangle  $, the first dashed orange box represents the entangling quantum gate $\widehat{\mathbf{J}}$ (or the judge). Both dark green small boxes represent  Alice's and Bob's quantum operator strategies labeled by $\widehat{\mathbf{U}}_{\text{A}} \left(  \theta_{\text{A}} ,  \phi_{\text{A}}   \right)$ and $\widehat{\mathbf{U}}_{\text{B}} \left(  \theta_{\text{B}} ,  \phi_{\text{B}}   \right)$, respectively. The second dashed orange box represents the unentangling quantum gate $\widehat{\mathbf{J}}^{\dagger} $, and finally a measurement operator for  both players.} \label{fig:QuantumCircuit}
\end{figure}

\textit{Quantum strategy operator.-} The definition of the strategy operator in quantum mechanics {(see Complement C$_{\text{II}}$ on pag. 176 of Ref. \cite{cohen1977Book} and also Eq. (3) of Ref. \cite{eisert1999})} is established by the {unitary} operator $\widehat{\mathbf{U}}\left( \theta ,\phi \right)$ {for one player (or one qubit)}  as follows:
\begin{equation}
\widehat{\mathbf{U}}\left( \theta ,\phi \right) =\left[ 
\begin{array}{cc}
\exp \left[ i\phi \right] \cos \frac{\theta }{2} & \sin \frac{\theta }{2} \\ 
-\sin \frac{\theta }{2} & \exp \left[ -i\phi \right] \cos \frac{\theta }{2}%
\end{array}%
\right]  \text{,} \label{OperadorEstrategiaGeral}
\end{equation}
with parameters $0\leq \theta \leq \pi $ and $0\leq \phi \leq \frac{\pi }{2}$. Opera and Television classical strategies have their quantum counterpart choosing $\left\{ \theta =0;\phi =0\right\} $ and $\left\{ \theta =\pi ;\phi = 0 \right\} $, respectively. In matrix notation, we have:
\begin{equation}
\widehat{\mathbf{O}}\equiv \widehat{\mathbf{U}}
\left( 0,0\right) =\left[ 
\begin{array}{cc}
1 & 0 \\ 
0 & 1
\end{array}
\right] \quad \widehat{\mathbf{T}}\equiv \widehat{\mathbf{U}}\left( \pi
,0\right) =\left[ 
\begin{array}{cc}
0 & 1 \\ 
-1 & 0
\end{array}
\right]  \text{.}
\end{equation}
For this two-player game, Alice can choose the Opera or Television quantum strategy operators defined by the following operators $\widehat{\mathbf{O}}_{\text{A}} \equiv \widehat{\mathbf{O}} \otimes \widehat{\mathbf{1}}$ or  $\widehat{ \mathbf{T}}_{\text{A}} \equiv \widehat{\mathbf{T}} \otimes \widehat{\mathbf{1}}$, while Bob's operators are $\widehat{\mathbf{O}}_{\text{B}}\equiv \widehat{\mathbf{1}} \otimes \widehat{\mathbf{O}}$ or $\widehat{\mathbf{T}}_{\text{B}}\equiv \widehat{\mathbf{1}}\otimes \widehat{\mathbf{T}}$. {In general, for any quantum strategy operator in the two player representation, Alice's operators are denoted by  $\widehat{\mathbf{U}}_{\text{A}} \left(  \theta_{\text{A}} ,  \phi_{\text{A}}   \right)  = \widehat{\mathbf{U}}  \left(  \theta_{\text{A}} ,  \phi_{\text{A}}    \right)  \otimes \widehat{\mathbf{1}}  $  where the Kronecker product is applied between the single qubit operator $\widehat{\mathbf{U}}   \left(  \theta  ,   \phi     \right)  $  and the identity operator  $\widehat{\mathbf{1}}$. Analogously, Bob's quantum strategy operators are denoted by $\widehat{\mathbf{U}}_{\text{B}}  \left(   \theta_{\text{B}} ,   \phi_{\text{B}}    \right) =  \widehat{\mathbf{1}} \otimes \widehat{\mathbf{U}}  \left(  \theta_{\text{B}} ,   \phi_{\text{B}}    \right)   $.}

\textit{The Judge operator.-} A two-qubit unitary, symmetric and invertible operator. The primary purpose of this gate is to entangle the initial quantum state, which carries the information of choice for both players. The judge operator is defined by:
\begin{equation}
\widehat{\mathbf{J}}\left( \lambda   \right)=\exp \left[ \frac{i \lambda }{2}\widehat{\mathbf{T}} \otimes \widehat{\mathbf{T}}\right] \text{,\qquad } \lambda \in \left[ 0,\frac{ \pi }{2}\right] \text{,}  \label{JuizForward}
\end{equation}
in which $ \lambda $ represents the level of entanglement, being $ \lambda = 0$ the absence of entanglement and  $\lambda = \frac{\pi }{2}$ maximum entanglement.

We briefly describe the Eisert-Wilkens-Lewenstein protocol \cite{eisert1999} to analyze the quantum battle of the sexes, and  Fig. \ref{fig:QuantumCircuit} summarizes the algorithm. The protocol starts at the quantum state  $\left\vert O_{\text{A}},O_{\text{B}}\right\rangle $.  Next, the entangled operator transforms the initial quantum state to produce $ \left\vert \psi _{0}\right\rangle =\widehat{\mathbf{J}}\left( \lambda   \right)\left\vert O_{\text{A}},O_{\text{B}}\right\rangle $. Each player acts locally performing their quantum strategy operators $\widehat{\mathbf{U}} _{\text{A} }\left( \theta _{\text{A} },\phi _{\text{A}}\right)   \widehat{\mathbf{U}} _{\text{B} }\left( \theta _{\text{B }},\phi _{\text{B}}\right) \left\vert \psi _{0}\right\rangle $. Finally, we apply the disentangling operator  $\widehat{\mathbf{J}}^{\dag }\left( \lambda   \right) $  providing the final quantum state $\left\vert \psi _{f}\right\rangle =\widehat{\mathbf{J}}^{\dag}\left( \lambda   \right)\widehat{\mathbf{U}} _{\text{A} }\left( \theta _{\text{A} },\phi _{\text{A}}\right)   \widehat{\mathbf{U}} _{\text{B} } \left( \theta _{\text{B }},\phi _{\text{B}}\right) \left\vert \psi _{0}\right\rangle $ at the end of the quantum circuit. {The study developed in this manuscript is performed in the highest regime of entanglement, $\lambda =  \frac{\pi}{2}$. Therefore, } the probability amplitudes of the ket representing the final quantum state are given by  Eq. (\ref{FinalQuantumState}):
\begin{equation}
\left\vert \psi _{f}\right\rangle =\left[ 
\begin{array}{c}
\cos \frac{\theta _{\text{A}}}{2}\cos \frac{\theta _{\text{B}}}{2}\cos
\left( \phi _{\text{A}}+\phi _{\text{B}}\right)  \\ 
\cos \frac{\theta _{\text{B}}}{2}\sin \frac{\theta _{\text{A}}}{2}\sin \phi
_{\text{B}}-\cos \frac{\theta _{\text{A}}}{2}\sin \frac{\theta _{\text{B}}}{2%
}\cos \phi _{\text{A}} \\ 
\cos \frac{\theta _{\text{A}}}{2}\sin \frac{\theta _{\text{B}}}{2}\sin \phi
_{\text{A}}-\cos \frac{\theta _{\text{B}}}{2}\sin \frac{\theta _{\text{A}}}{2%
}\cos \phi _{\text{B}} \\ 
\sin \frac{\theta _{\text{A}}}{2}\sin \frac{\theta _{\text{B}}}{2}+\cos 
\frac{\theta _{\text{A}}}{2}\cos \frac{\theta _{\text{B}}}{2}\sin \left(
\phi _{\text{A}}+\phi _{\text{B}}\right) 
\end{array}%
\right] \label{FinalQuantumState}
\end{equation}

\begin{SCfigure}
    $\begin{array}{c}
     \mbox{ (a) } \\ [-0.00cm]
     \epsfxsize=3.0in      \epsffile{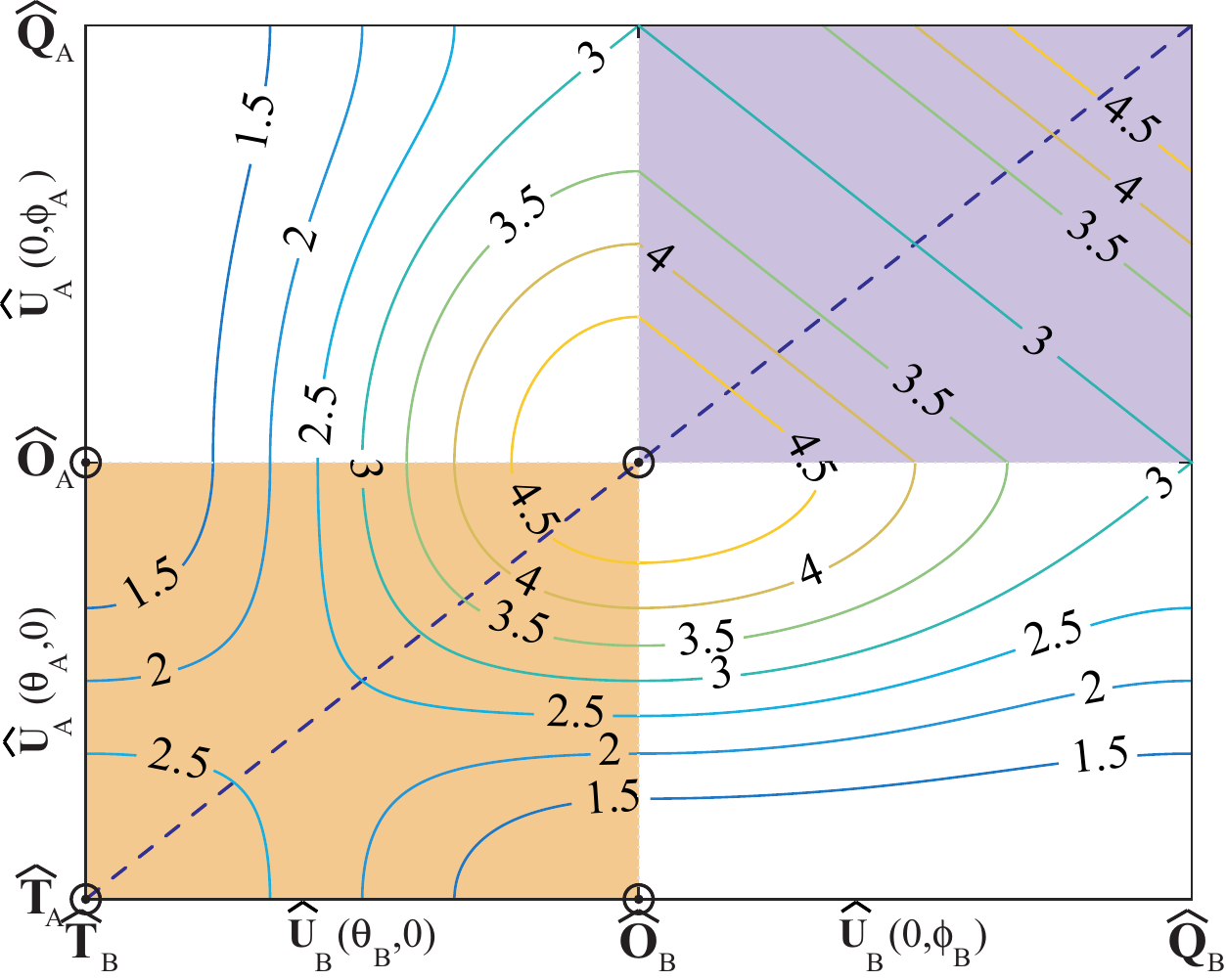} \\ [0.0cm]
     \mbox{ (b) }  \\ [0.0cm]
     \epsfxsize=3.0in      \epsffile{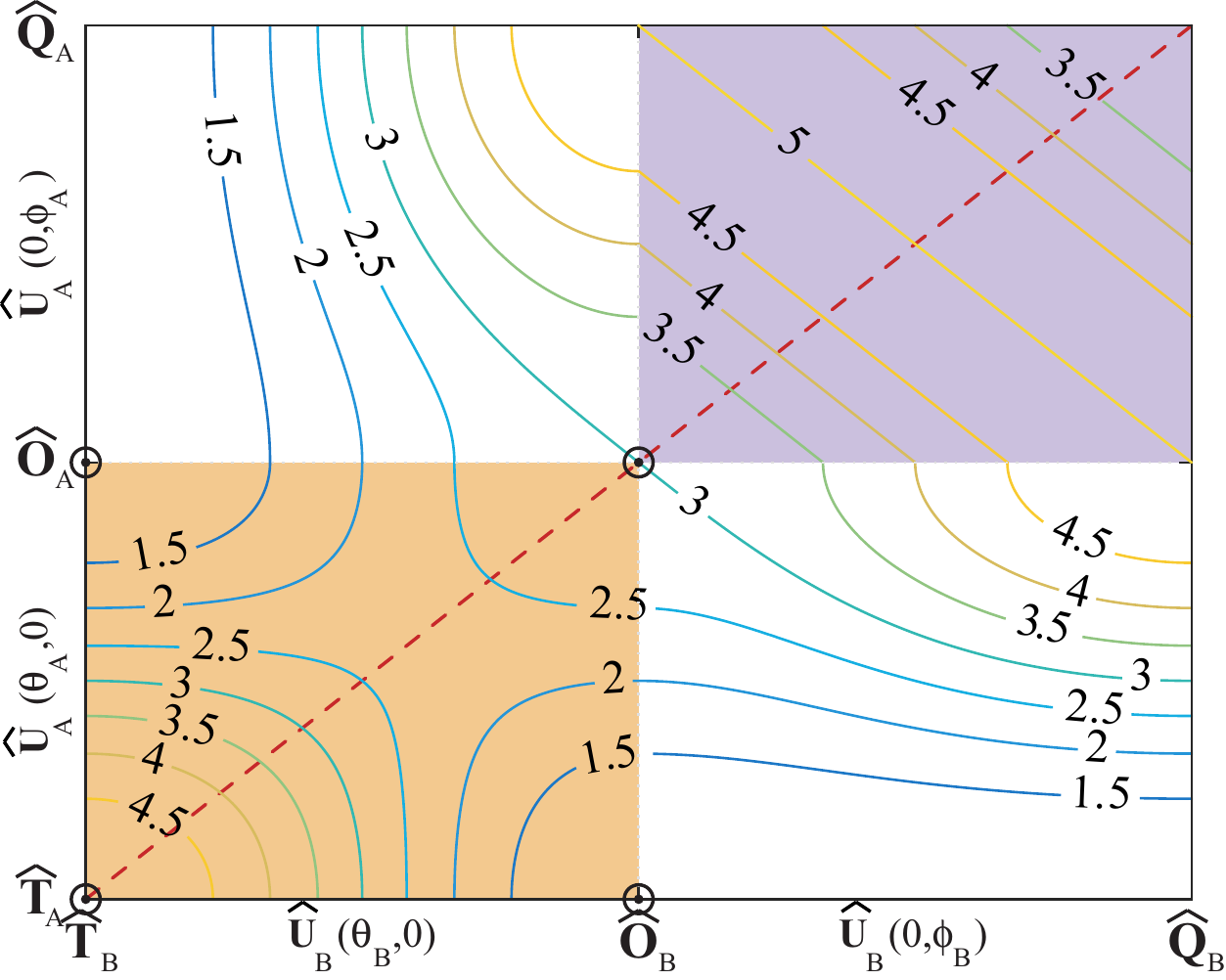} \\ [0.0cm]
     \mbox{ (c) }  \\ [0.0cm]
     \epsfxsize=3.0in      \epsffile{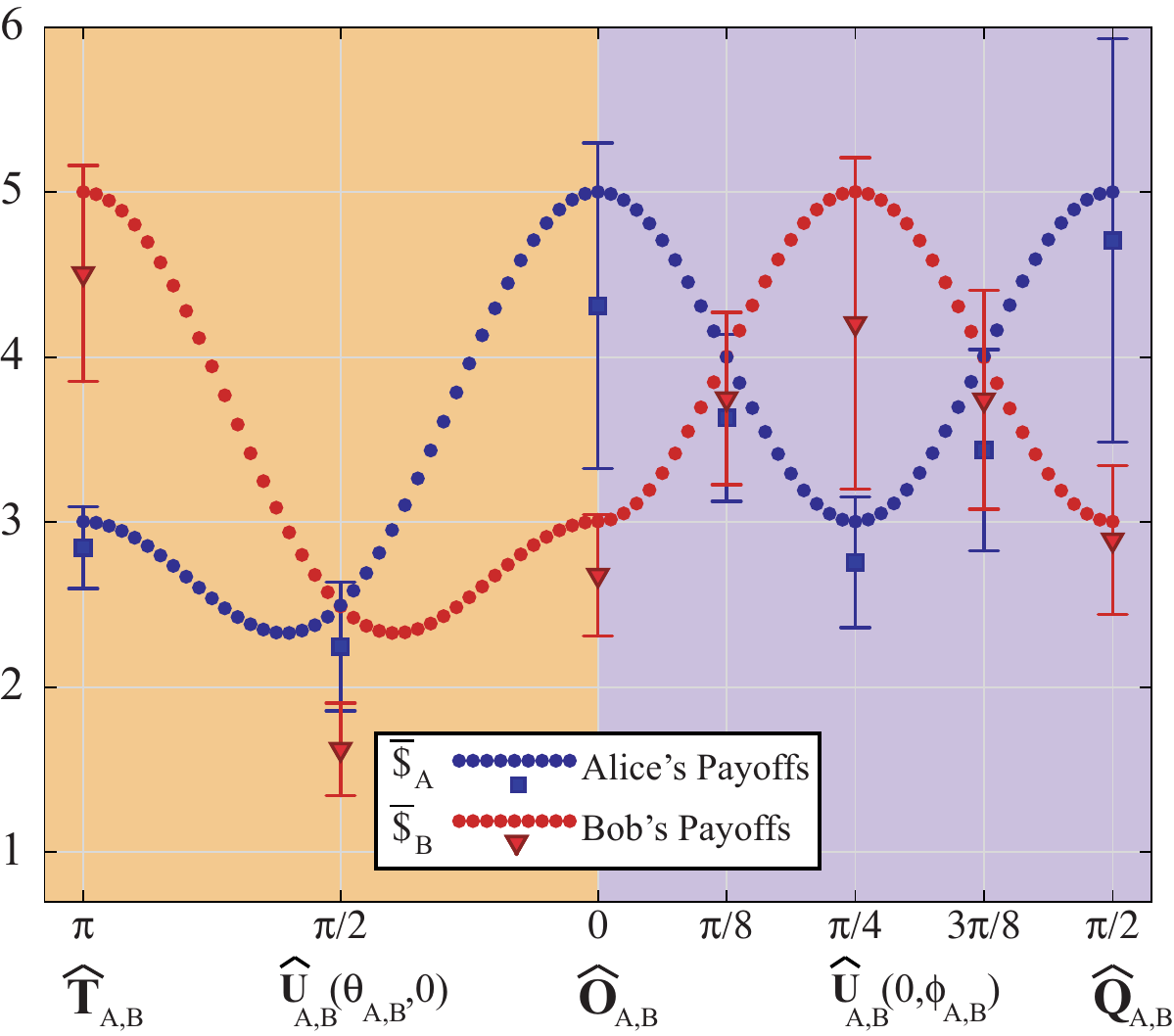}  
     \end{array}$
\caption{ (Color online) Payoffs for the battle of the sexes quantum game {as a function of the parameter domain for the quantum strategy operators}.  The parameter domain is divided into two parts: the first one satisfies   $  \theta_{\text{A}}, \theta_{\text{B}} \in \left[ 0 ,  \pi  \right]  $ and   $\phi_{\text{A,B}} = 0 $, which corresponds to moving from  Television $\widehat{\mathbf{T}}_{\text{A,B}} = \widehat{\mathbf{U}}_{\text{A,B}}\left( \theta_{\text{A,B}}=\pi ,\phi_{\text{A,B}} = 0 \right) $ to Opera  $\widehat{\mathbf{O}}_{\text{A,B}} = \widehat{\mathbf{U}}_{\text{A,B}}\left( \theta_{\text{A,B}}=0 ,\phi_{\text{A,B}} = 0 \right)$ strategy operator; the second one satisfies  $\theta_{\text{A,B}} = 0 $ and  $   \phi_{\text{A}} , \phi_{\text{B}} \in  \left[ 0 ,  \frac{\pi}{2}  \right]      $,  which corresponds to moving from the Opera $\widehat{\mathbf{O}}_{\text{A,B}} = \widehat{\mathbf{U}}_{\text{A,B}}\left( \theta_{\text{A,B}}=0 ,\phi_{\text{A,B}} = 0 \right)$ strategy to the $\widehat{\mathbf{Q}}_{\text{A,B}} = \widehat{\mathbf{U}}_{\text{A,B}}\left( \theta_{\text{A,B}}=0 ,\phi_{\text{A,B}} = \frac{\pi}{2} \right)$ strategy operators, {where $\alpha=5$,   $\beta=3$ and  $\gamma=1$  obeying the relation $\alpha > \beta > \gamma $.} {In Tab. \ref{tab:PayoffsBattleSexesNovo} are resumed the payoffs values related to the most important quantum strategy operators $\widehat{\mathbf{T}}_{\text{A,B}}$,  $\widehat{\mathbf{O}}_{\text{A,B}}$,  $\widehat{\mathbf{Q}}_{\text{A,B}}$.}  (a) Contour graph representing Alice's   payoffs, $\overline{\$}_{\text{A}}$.   (b) Contour graph representing Bob's   payoffs, $ \overline{\$}_{\text{B}}$.  The small black circles in Fig. \ref{fig:PayoffBattleOfSexes}(a) and  Fig. \ref{fig:PayoffBattleOfSexes}(b) correspond to the classical counterpart.  (c) Cross-section represented by the red and blue dashed line of Fig. \ref{fig:PayoffBattleOfSexes}(a) and Fig.  \ref{fig:PayoffBattleOfSexes}(b), indicating the payoff values of Alice and Bob, when both use the same quantum strategy operators.  Blue dots and squares represent theoretical prediction and experimental results of Alice's payoffs, respectively. Red dots and triangles represent theoretical prediction and experimental results of Bob's payoffs, respectively. {This figure is a dimensionally reduced representation of the payoffs constrained to a single plane. In Fig. \ref{fig:ManyAngularParametersPayoffs} we  sketch a few other planes of interest based on the general description made in appendix \ref{sec:TheoreticalAnalysis}.} } \label{fig:PayoffBattleOfSexes}
\end{SCfigure}

To perform the analysis of the game, we compute the probability values represented by $ P_{s_{\text{A}},s_{\text{B}}}=\left\vert \left\langle s_{\text{A}},s_{\text{B }}|\psi _{f}\right\rangle \right\vert ^{2}$, where $s_{\text{A,B}}\in \left\{ O,T\right\} $ and then we evaluate the payoff, similarly to equation 2 of reference  \cite{eisert1999}, for both players {(see Eq.(\ref{POOWithEntangledParameters}-\ref{PTTWithEntangledParameters}) of appendix \ref{sec:TheoreticalAnalysis} for an explicit representation of the probabilities)}. Using the bi-matrix detailed in Tab. \ref{tab:PayoffsBattleSexes}, we can achieve the expressions of the payoff in terms of $\alpha$, $\beta$ and $\gamma$:
\begin{eqnarray}
\overline{\$}_{\text{A}} \left( \theta _{\text{A,B}},\phi _{\text{A,B}} \right)  &=&\alpha P_{OO}+\gamma  \left(  P_{OT} + P_{TO}  \right)  +\beta P_{TT}\text{,} \quad \label{PayoffAlice} \\
\overline{\$}_{\text{B}} \left( \theta _{\text{A,B}},\phi _{\text{A,B}} \right)  &=&\beta P_{OO}+\gamma  \left(  P_{OT} + P_{TO}  \right) +\alpha P_{TT}\text{.} \quad \label{PayoffBob}
\end{eqnarray}
These expressions depend on four angular parameters, $    \theta _{\text{A,B}},\phi _{\text{A,B}}   $. To give a visual description of the game, with the values  $\alpha=5$, $\beta=3$ and $\gamma=1$, we make a contour plot in  Fig. \ref{fig:PayoffBattleOfSexes}(a) and Fig. \ref{fig:PayoffBattleOfSexes}(b). The figures describe the payoff values that Alice and Bob receive in response to their performed strategy. The horizontal (vertical) axis of the plot corresponds to the quantum strategy operators performed by Bob (Alice).  Both axes start at $ \widehat{\mathbf{T}}_{\text{A,B}}
 = \widehat{\mathbf{U}}_{\text{A,B}}\left( \theta_{\text{A,B}} = \pi ,\phi_{\text{A,B}} = 0 \right)   $  (Television) then $\theta_{\text{A,B}} $ moves from $\pi$ to 0 arriving at $ \widehat{\mathbf{O}}_{\text{A,B}}  = \widehat{\mathbf{U}}_{\text{A,B}}\left( \theta_{\text{A,B}} = 0 ,\phi_{\text{A,B}} = 0 \right) $ (Opera) (the dark orange region in Fig. \ref{fig:PayoffBattleOfSexes}). Past this point on the axis, the parameter $ \phi_{\text{A,B}}$ varies from 0 to $\frac{\pi}{2} $ while $\theta_{\text{A,B}}$ remains constant at zero, reaching the quantum strategy operator $ \widehat{\mathbf{Q}}_{\text{A,B}}   = \widehat{\mathbf{U}}_{\text{A,B}}\left( \theta_{\text{A,B}} = 0 ,\phi_{\text{A,B}} = \frac{\pi}{2} \right) $  (the light purple  region in Fig. \ref{fig:PayoffBattleOfSexes}). In Fig. \ref{fig:PayoffBattleOfSexes}(c) we sketch the payoff of each player considering both are performing the same quantum strategy operator . The blue (red) dots represent the theoretical prediction for Alice's (Bob's) payoffs. Also, square (triangle) symbols represent the experimental results of Alice's (Bob's) payoffs computed from {experimental quantum states as detailed in Fig.  \ref{fig:DensityMatrices} (we will present explanations about the experimental setup in the next section)}.

\begin{table}[!ht]
\begin{center}
\begin{tabular}{lcccccc} \hline\hline
\multicolumn{2}{c}{ }																&           \multicolumn{5}{c}{Bob}  \\ 
\multicolumn{2}{c}{ }																&	$ \widehat{\mathbf{T}}_{\text{B}}$		& 	\quad	$ \widehat{\mathbf{O}}_{\text{B}}$	\quad	&  $\widehat{\mathbf{U}}_{\text{B}}\left( 0, \frac{ \pi }{8} \right) $	& 	$\widehat{\mathbf{U}}_{\text{B}}\left( 0, \frac{3 \pi }{8} \right) $ 	 &	$\widehat{\mathbf{Q}}_{\text{B}} $ \\  [3pt]   \hline
 		&					$ \widehat{\mathbf{T}}_{\text{A}}$							&	$ \left( 3, 5 \right)$	& 	\quad	$ \left( 1 , 1 \right)$		\quad	& 		$ \left( 1   ,  1	\right)$	         		& 	$ \left( 1   ,  1	\right)$	        				& 	$ \left( 1   ,  1	\right)$			\\
		&					$ \widehat{\mathbf{O}}_{\text{A}}$							&	$ \left( 1, 1 \right)$	& 	\quad	$ \left( 5, 3 \right)$		\quad	& 		$ \left( 4.7 , 3.3	\right)$					& 	$ \left( 3.3 , 4.7	\right)$							 &	$ \left( 3 ,  5	\right)$			\\
 Alice	&		$\widehat{\mathbf{U}}_{\text{A}}\left( 0, \frac{ \pi }{8} \right) $		&	$ \left( 1, 1 \right)$	& 	\quad	$ \left( 4.7 , 3.3 \right)$	\quad	& 		$ \left( 4   ,  4	\right)$					& 	$ \left( 3   ,  5	\right)$							 &	$ \left( 3.3   ,  4	.7\right)$			\\ 
		&		$\widehat{\mathbf{U}}_{\text{A}}\left( 0, \frac{3\pi }{8} \right) $		&	$ \left( 1, 1 \right)$	& 	\quad	$ \left( 3.3 , 4.7 \right)$	\quad	& 		$ \left( 3   ,  5	\right)$					& 	$ \left( 4   ,  4	\right)$							 &	$ \left( 4.7   ,  3.3\right)$			\\
 		&					$ \widehat{\mathbf{Q}}_{\text{A}}$							&	$ \left( 1, 1 \right)$	& 	\quad	$ \left( 3 , 5 \right)$		\quad	& 		$ \left( 3.3   ,  4.7	\right)$	         		& 	$ \left( 4.7   ,  3.3	\right)$	        				& 	$ \left( 5   ,  3	\right)$	 \\ [3pt] \hline \hline
\end{tabular}
\caption{Bi-matrix of payoffs  for the battle of the sexes game  \cite{marinatto2000}. The first (second) entry in the parenthesis denotes Alice's (Bob's) payoffs. The payoffs values obey $ \alpha > \beta > \gamma $ with  $\alpha=5$, $\beta=3$ and $\gamma=1$. The Television and Opera  strategies are defined following the  quantum strategy operators   $ \widehat{\mathbf{T}}_{\text{A,B}} = \widehat{\mathbf{U}}_{\text{A,B}}\left(  \pi  ,    0  \right) $  and   $ \widehat{\mathbf{O}}_{\text{A,B}} = \widehat{\mathbf{U}}_{\text{A,B}}\left(  0  ,    0  \right) $, respectively.   The quantum strategy operators which satisfy the Marinatto-Weber criteria are   $\widehat{\mathbf{U}}_{\text{A,B}}\left( 0 ,  \frac{ \pi }{8}  \right) $  and   $\widehat{\mathbf{U}}_{\text{A,B}}\left( 0 ,  \frac{3\pi }{8}  \right) $. The operator $\widehat{\mathbf{Q}}_{\text{A,B}} = \widehat{\mathbf{U}}_{\text{A,B}}\left( 0, \frac{\pi }{2} \right)  $. This table summarizes the data shown in Fig. \ref{fig:PayoffBattleOfSexes}, which correspond to the most relevant quantum strategy operators and their respective payoff values for both players. }	\label{tab:PayoffsBattleSexesNovo}
\end{center}
\end{table}

\section{Experimental procedures} \label{sec:ExperimentalDetails}

In this Nuclear Magnetic Resonance (NMR) experiment, we apply a similar procedure of the first implementation of the prisoner's dilemma  \cite{du2002}. Here, we describe the most important details. The experiment is performed on a Tecmag/Jastec 400 MHz spectrometer at room temperature (20 $^{\circ}$C). An enriched Chloroform sample ($^{13}$CHCl$_{3}$) was used as a two-qubit spin system, in which the Hydrogen (Carbon) nuclei carries the information about the quantum strategy operator performed by Alice (Bob). The stoichiometry of the sample is the dissolution of 12 \% enriched chloroform and 88 \% deuterated acetone for a total volume of 600 $\mu$L. We then seal the sample on a standard 5 mm NMR tube. The {dual channel} NMR probe head is a  VARIAN 10 mm for solution samples. {The $\pi/2$-pulse time is calibrated at 49 $\mu$s, for both channels. Also, the transversal and longitudinal relaxation times are $T_{2}^{\text{H}} \approx 1.4 $ s and  $T_{1}^{\text{H}} \approx 11.74  $ s for Hydrogen nuclei and    $T_{2}^{\text{C}} \approx 1.2 $ s and  $T_{1}^{\text{C}} \approx 17.11  $ s for Carbon  nuclei, respectively. The recycle delay time is 120 s}.

The Hamiltonian of the two-spin system (Hydrogen and Carbon nuclei in this case) is the combination of three energy contributions \cite{oliveira2007}: the first one is the Zeeman interaction, which is the interaction between the magnetic moment of the nuclei with a strong static magnetic field along the $z$-axis $\mathbf{B}_{0}=B_{0}\mathbf{e}_{z}$, the second one is the $J$-coupling between neighbor interacting nuclear spins, and the third one is the interaction between the magnetic moment of the nuclei with an external time-dependent weak magnetic field parallel to the $xy$-plane. Those energy contributions are represented in the rotating frame description by the Hamiltonian:
\begin{eqnarray}
\hat{\mathcal{H}} & = & -\left( \omega ^{\text{H}}_{L}-\omega _{rf}^{\text{H} }\right) \hat{\mathbf{I}}_{z}^{\text{H}}-\left( \omega ^{\text{C} }_{L}-\omega _{rf}^{\text{C}}\right) \hat{\mathbf{I}}_{z}^{\text{C}}+2\pi J \hat{\mathbf{I}}_{z}^{\text{H}}\hat{\mathbf{I}}_{z}^{\text{C}} \nonumber \\ 
					& & + \omega _{1}^{\text{H}}\left( \hat{\mathbf{I}}_{x}^{\text{H}} \cos \phi ^{\text{H}}+\hat{\mathbf{I}}_{y}^{\text{H}}\sin \phi ^{\text{H}}\right) + \omega _{1} ^ {\text{C}}\left( \hat{\mathbf{I}}_{x}^{\text{C}}\cos \phi ^{ \text{C} } + \hat{\mathbf{I}} _ {y} ^ {\text{C}} \sin \phi ^{\text{C}} \right) \text{,}
\end{eqnarray}
in which $\omega ^{\text{H,C}}_{L}$ are the Larmor frequencies of each nuclear specie, $\omega _{rf}^{\text{H,C} } $ are the radio-frequencies of the external time-dependent weak magnetic field, $\omega _{1}^{\text{H,C}}$ are the intensity of the external time-dependent weak magnetic field, $\phi ^{\text{H,C}}$ define the direction of the external time-dependent weak magnetic field and $\hat{\mathbf{I}}_{x,y,z}^{\text{H,C}}$ are the spin angular momentum operators.

\begin{figure}[ht!]
\begin{center}
\includegraphics[width=0.80\textwidth]{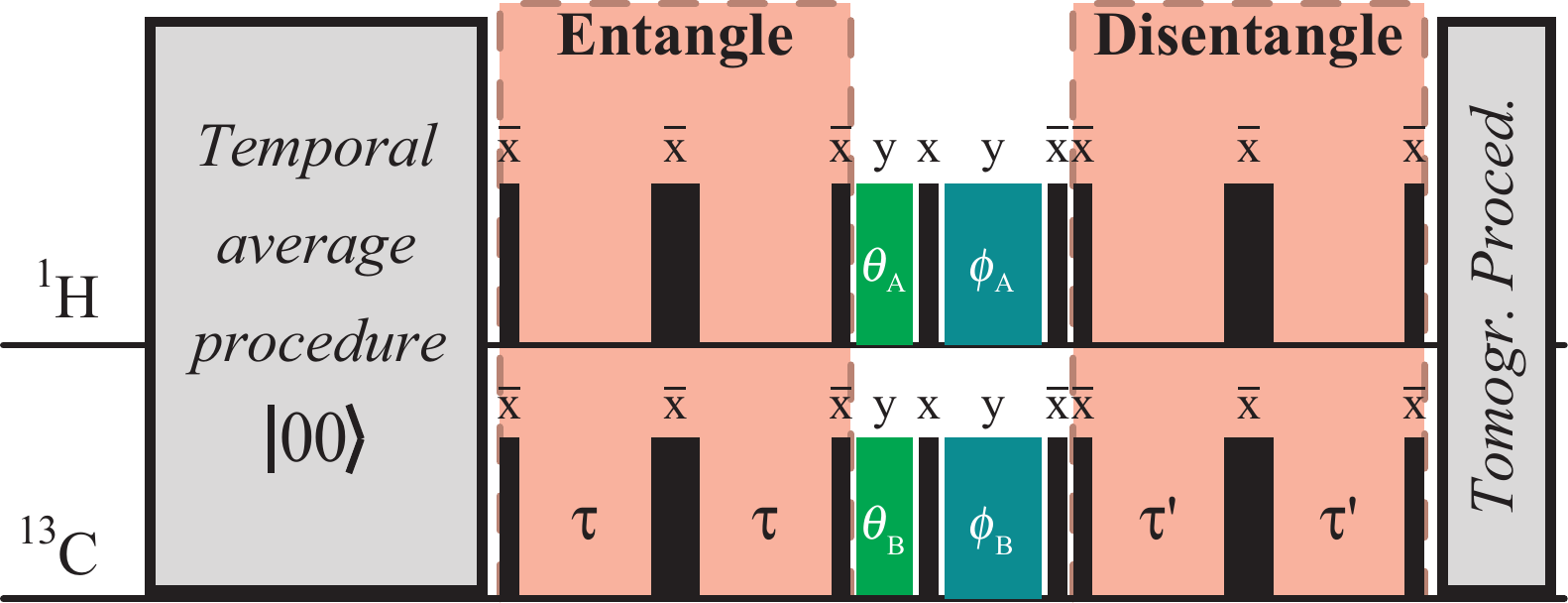}
\end{center}
\caption{(Color online)  NMR Pulse sequence to implement and to monitor the Eisert-Wilkens-Lewenstein protocol \cite{eisert1999}. The thicker black bars represent $\pi$-pulses, the thinner black bars represent $\pi/2$-pulses. Light and dark green bars represent variable length pulses and depend on the angular parameters $\theta_{\text{A,B}}$ and $\phi_{\text{A,B}}$, and we use them to control the quantum game. Above each bar, the letters $x$ and $y$ denote the positive direction of the axis to perform the radio-frequency pulse, and $\overline{\text{x}}$ and $\overline{\text{y}}$ denote the negative direction of that axis to perform the radio-frequency pulse. The gaps between the bars indicate free evolutions.} \label{fig:PulseSequence}
\end{figure}

The main task of the experimental setup is to perform the game protocol depicted in Fig. \ref{fig:QuantumCircuit}. For that purpose, the game protocol is encoded as a pulse sequence (Fig. \ref{fig:PulseSequence}), and here we introduce the most important details.

\textit{Initialization.-} Standard high temperature NMR description of the quantum state is expressed as a first order expansion of the density matrix definition
\[
\hat{\boldsymbol{\rho }}=\frac{1}{\mathcal{Z}}\exp \left[ -\beta \hat{ \mathcal{H}}_{0}\right] \simeq \frac{1}{\mathcal{Z}}\hat{\mathbf{1}} _{4\times 4}-\frac{\beta}{\mathcal{Z}} \hat{\mathcal{H}}_{0}\text{,}
\]
where $\mathcal{Z}$ is the partition function, $\beta =\left( k_{B}T\right) ^{-1}$, $k_{B}$ is the Boltzmann constant and $T$ the room temperature, $\hat{\mathcal{H}}_{0}$ is the Zeeman Hamiltonian at the laboratory frame
$
\hat{\mathcal{H}}_{0}=- \hbar \omega _{L}^{\text{H}}\hat{\mathbf{I}}_{z}^{\text{H} } - \hbar \omega _{L}^{\text{C}}\hat{\mathbf{I}}_{z}^{\text{C}}\text{,}
$ so that the density matrix is expressed as
\begin{equation}
\hat{\boldsymbol{\rho }}\simeq \frac{1}{\mathcal{Z}}\hat{\mathbf{1}} _ {4\times 4} + \frac{ \beta \hbar \omega _{L}^{\text{H}}}{\mathcal{Z}}\left( \hat{\mathbf{I}}_{z}^{\text{H}}+\frac{\omega _{L}^{\text{C}}}{\omega _{L} ^ { \text{H} } } \hat{\mathbf{I}}_{z}^{\text{C}}\right) \text{.}
\end{equation}
The main purpose of the initialization procedure is to transform the second term of the expanded density matrix into a contribution with equivalent properties of an effective pure state. In order to do that, we use the temporal average procedure, which consists of the permutation of populations of the density matrix to reduce the noise level \cite{knill1998}. As the density operator is represented by a $4\times 4$ matrix, it represents four populations  labelled $p_{1}$, $p_{2}$, $p_{3}$, and $p_{4}$. The procedure involves three stages: in the first one, we perform no  permutation of populations; in the second one there is a permutation of populations as $p_{2}\rightarrow p_{4}^{\prime }$ , $p_{3}\rightarrow p_{2}^{\prime }$ , and $p_{4}\rightarrow p_{3}^{\prime }$ (see Eq. (6) of Ref. \cite{knill1998}); in the third step we perform the inverse permutation of populations as $p_{4}\rightarrow p_{2}^{\prime \prime }$ , $ p_{3}\rightarrow p_{4}^{\prime \prime }$ , and $p_{2}\rightarrow
p_{3}^{\prime \prime }$ (see Eq. (7) of Ref. \cite{knill1998}).\ The average density matrix is represented as
\begin{equation}
\hat{\boldsymbol{\rho }}\simeq \left( \frac{1}{\mathcal{Z}}-\frac{\epsilon }{ 4 } \right) \hat{ \mathbf{1} } _ { 4 \times 4 } + \epsilon \left\vert 00\right\rangle \left\langle 00\right\vert \text{,}
\end{equation}
where $\epsilon =\frac{4\beta \hbar }{3\mathcal{Z}} \frac{ \omega _ {L} ^ {\text{ H } } + \omega _ {L} ^ {\text{C}} } { 2 } \sim 1.34\times 10^{-5}$.

Therefore, the quantum system is initialized at the quantum state $\left\vert 0,0\right\rangle \equiv \left\vert O_{\text{A}},O_{\text{B}}\right\rangle $  and experimentally achieved using the temporal average procedure \cite{knill1998}.

\textit{The Judge operator.-} The action of the Judge in the game protocol is to entangle and to disentangle the quantum state of the two-qubit spin system,  as discussed in Ref.  \cite{du2002}. Here, we represent both procedures by the orange square on Fig. \ref{fig:PulseSequence}. Both actions are performed using three radio frequency pulses and two free evolutions. The entangling procedure starts performing a $\pi$/2-pulse along the negative $x$-axis on both channels; next, a free evolution during  $ \tau = \frac{\lambda}{2 \pi J} $ is executed;  we apply a  $\pi$-pulse along the negative $x$-axis on both channels, another free evolution during  $ \tau = \frac{\lambda}{2 \pi J} $ and finally we apply another $\pi$/2-pulse along the negative $x$-axis on both channels. The disentangling procedure is very similar to the previous one, the only difference being the free evolution occurs during   $ \tau^{\prime} = \frac{2\pi-\lambda}{2 \pi J} $. The control parameter of the Judge's action is established by  $ \lambda  \in \left[  0 , \frac{\pi}{2}\right]$ and it is adjusted to $\frac{\pi}{2}$ to emulate  the regime of maximum entanglement of the quantum states, and the parameter $J = 215$ Hz is the scalar coupling constant between the $^{1}$H and $^{13}$C nuclei.

\textit{Quantum strategies operators.-} The theoretical representation of strategy operator is described in Eq. (\ref{OperadorEstrategiaGeral}). It depends on two angular parameters $ \theta $  and $\phi$. The $\theta$-control is performed by a radio-frequency pulse along the positive  $y$-axis. The  $\phi$-control is performed as a composite $z$-pulse sequence \cite{freeman1981} by three radio-frequency pulses as follows:  a $\pi$/2-pulse along the positive $x$-axis, $\phi$-pulse along the positive $y$-axis{, performed twice,} and   a $\pi$/2-pulse along the negative $x$-axis.  This description is for both channels, so that the   $\theta$- and  $\phi$-controls are represented in Fig. \ref{fig:PulseSequence} by light and dark green squares, respectively.

\textit{Tomography procedure.-} This is a read out procedure used to reconstruct the density matrix. The procedure   is performed by a series of nine experiments described as {II, IX, IY, XI, XX, XY, YI, YX, YY}, in which the first (second) entry represents an action on the Hydrogen (Carbon) nuclei \cite{cory1997,chuang1998PRSLA,teles2007}. The label described ``I'' means no radio-frequency pulse, ``X'' represents to perform a $\pi/2$-pulse along the positive $x$-axis, and  ``Y'' means performing a $\pi/2$-pulse along the positive $y$-axis. After performing the respective radio-frequency pulses, the detector is turned on to observe the free induction decay.

\begin{SCfigure}
$
\begin{array}{c@{\hspace{0.030in}}c@{\hspace{0.030in}}c}
\begin{array}{c}
\mbox{ {   	} }	\\ [-21.0cm]
\mbox{ { $\theta_{\text{A,B}}=\pi $	} }	\\ [0.0cm]
\mbox{ {   $ \phi_{\text{A,B}}=0$	} }	\\ [0.0cm]
\mbox{ {          $F=0.9709$			} }	\\ [1.7cm]
\mbox{ { $\theta_{\text{A,B}}=\frac{\pi}{2}$	} }	\\ [0.0cm]
\mbox{ {   $ \phi_{\text{A,B}}=0$			} }	\\ [0.0cm]
\mbox{ {          $F=0.9306$					} }	\\ [1.7cm]
\mbox{ { $\theta_{\text{A,B}}=0$		} }	\\ [0.0cm]
\mbox{ {   $ \phi_{\text{A,B}}=0$	} }	\\ [0.0cm]
\mbox{ {          $F=0.9542$			} }	\\ [1.7cm]
\mbox{ { $\theta_{\text{A,B}}=0$				} }	\\ [0.0cm]
\mbox{ { $\phi_{\text{A,B}}=\frac{\pi}{8}$	} }	\\ [0.0cm]
\mbox{ {          $F=0.9652$					} }	\\ [1.7cm]
\mbox{ { $\theta_{\text{A,B}}=0$				} }	\\ [0.0cm]
\mbox{ { $\phi_{\text{A,B}}=\frac{\pi}{4}$	} }	\\ [0.0cm]
\mbox{ {          $F=0.9522$					} }	\\ [1.7cm]
\mbox{ { $\theta_{\text{A,B}}=0$				} }	\\ [0.0cm]
\mbox{ { $\phi_{\text{A,B}}=\frac{3\pi}{8}$	} }	\\ [0.0cm]
\mbox{ {          $F=0.9557$					} }	\\ [1.7cm]
\mbox{ { $\theta_{\text{A,B}}=0$				} }	\\ [0.0cm]
\mbox{ { $\phi_{\text{A,B}}=\frac{\pi}{2}$	} }	\\ [0.0cm]
\mbox{ {          $F=0.9480$					} } 
\end{array}	&  \epsfxsize=1.25in \epsffile{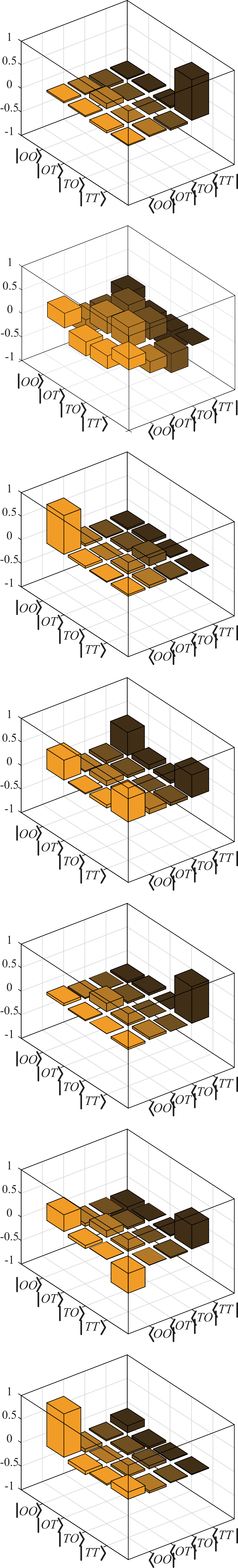}	&  \epsfxsize=1.25in \epsffile{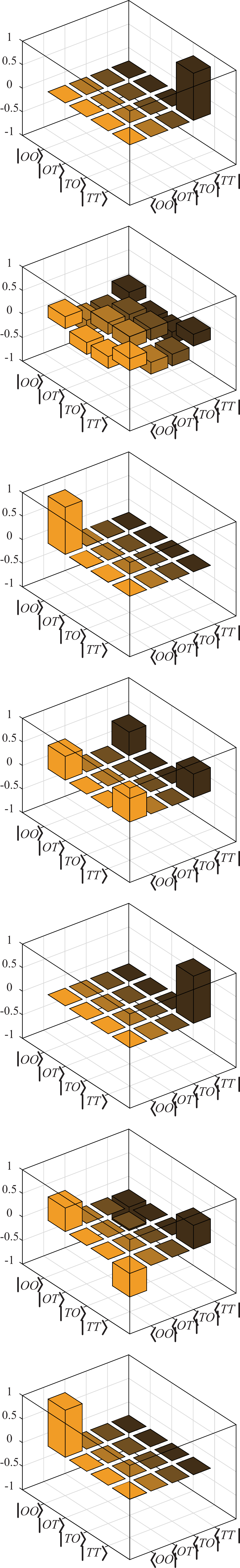}	 
\end{array}
$
\caption{ (Color online) Bar charts representing the real elements of the tomographed density matrices for the set of angular parameters used for the chosen quantum strategies operators. Imaginary elements are neglected as their values are lower than 0.1. The left (right)  column of the bar charts represents experimental (theoretical) density matrices. The fidelity  values $F$ of the experimental density matrices are above 93 \%. } \label{fig:DensityMatrices}
\end{SCfigure}

This procedure was used in other applications as detailed in Ref. \cite{auccaise2011A,auccaise2011B}. The result of the implementation of the tomography procedure using the experimental setup as described above can be sketched as bar charts on Fig. \ref{fig:DensityMatrices}. We plot the real part of the density matrices for seven sets of parameters that make up a quantum strategy operator each. We display on the  left (right) bar charts of Fig. \ref{fig:DensityMatrices} the experimental (theoretical) density matrices. Theoretical density matrices are computed from Eq. (\ref{FinalQuantumState}).  Fidelity values are above 93 \%.  Those seven density matrices are used to calculate the experimental data represented by red triangle and blue square symbols in Fig. \ref{fig:PayoffBattleOfSexes}(c). {The payoff values from the experimental data are computed using the diagonal elements of the density matrix displayed in Fig. 4, where the labels $ \left\vert OO\right\rangle \left\langle OO\right\vert  $  represents the probabilities $ P_{OO}  $, and similarly for the other elements of the density matrix and probabilities of Eq. (\ref{PayoffAlice}) and Eq. (\ref{PayoffBob}).} An error data analysis is displayed on Tab. \ref{tab:ExperimentalData1} in which we compute the error $\overline{\varepsilon  }_{\text{A,B}}$, between the theoretical expectation value $ \overline{\$}_{\text{A,B}}$ and the experimental results $\overline{\pounds}_{\text{A,B}}$ of Alice's and Bob's payoffs, represented by the expression $\overline{\varepsilon  }_{\text{A,B}} = \overline{\pounds}_{\text{A,B}} \times \left( 1 - F\right) $. Also, the error values are sketched in Fig. 2(c) of the main text as a bar error of experimental results represented by triangle and square symbols.

\begin{table}[!h!p!b]
\begin{center}
\caption{Theoretical and Experimental values of Alice's and Bob's payoffs for the main quantum strategies operators analyzed in this study. We use $ \overline{\pounds}_{\text{A,B}}$ to represent the payoffs computed from the experiments and $\overline{\varepsilon  }_{\text{A,B}} $ the error between theoretical prediction and experimental results. }
\begin{tabular}{ccccc} \hline  \hline  
Quan. Strat. 	&	  \multicolumn{2}{c}{Theoretical} 	&  \multicolumn{2}{c}{Experimental} \\ 
$\left( \theta_{\text{A}}, \phi_{\text{A}} ,  \theta_{\text{B}}, \phi_{\text{B}} \right)$	&	\quad 		$\overline{\$}_{\text{A}}$ \quad	& \quad	$ \overline{\$}_{\text{B}}$ \quad	& \quad	$ \overline{\pounds}_{\text{A}} \pm \overline{\varepsilon }_{\text{A}}  $ \quad 	& \quad	$ \overline{\pounds}_{\text{B}} \pm \overline{\varepsilon  }_{\text{B}}  $  \quad  	\\  \hline
$\left( \pi,0 , \pi,0  \right)$					&	3.0	&	5.0	&	$2.8	 	\pm	0.2 $	&	$4.5 	\pm	0.7 $		\\
$\left( \frac{ \pi }{2}, 0 , \frac{ \pi }{2}, 0 \right)$		&	2.5	&	2.5	&	$2.2 	\pm	0.4 $	&	$1.6 	\pm	0.3 $		\\
$\left( 0, 0 , 0, 0 \right)$					&	5.0	&	3.0	&	$4.3 	\pm	1.0 $	&	$2.7 	\pm	0.4 $		\\
$\left( 0, \frac{ \pi }{8} ,  0, \frac{ \pi }{8} \right)$		&	4.0	&	4.0	&	$3.6 	\pm	0.5 $	&	$3.7 	\pm	0.5 $		\\
$\left( 0, \frac{ \pi }{4} , 0, \frac{ \pi }{4} \right)$		&	3.0	&	5.0	&	$2.8 	\pm	0.4 $	&	$4.2 	\pm	1.0 $		\\
$\left( 0, \frac{3\pi }{8} ,  0, \frac{3\pi }{8} \right)$		&	4.0	&	4.0	&	$3.4 	\pm	0.6 $	&	$3.7 	\pm	0.7 $		\\
$\left( 0, \frac{ \pi }{2} ,  0, \frac{ \pi }{2} \right)$		&	5.0	&	3.0	&	$4.7 	\pm	1.2 $	&	$2.9 	\pm	0.5 $		\\ \hline  \hline 
\end{tabular}
\label{tab:ExperimentalData1}
\end{center}
\end{table}

\section{Discussion}\label{sec:Discussion}

To determine a satisfactory condition as the solution to the game, {we use the main point that the Marinatto-Weber protocol highlight, ``both players have the same degree of satisfaction" \cite{marinatto2000}. Therefore, the mathematical procedure compatible with that information is to compare  Alice's and Bob's payoffs. Thereby, }   we compare Fig. \ref{fig:PayoffBattleOfSexes}(a) and Fig. \ref{fig:PayoffBattleOfSexes}(b){, or analogously Eq. (\ref{PayoffAlice}) and Eq. (\ref{PayoffBob}) for an analytical procedure (an extended theoretical description  developed to calculate the values of the angular parameters which satisfy the Marinatto-Weber condition is shown in appendix \ref{sec:TheoreticalAnalysis})}. We can observe that the highest values of payoff with both players having equal  payoff  occurs when they play with quantum strategy operators defined by $\widehat{\mathbf{U}}_{\text{A}}\left( 0 ,\phi _{\text{A}}\right) $ and  $\widehat{\mathbf{U}}_{\text{B}}\left(  0 ,\phi _{\text{B}}\right) $. From   Eq.  (\ref{PayoffAlice}) and Eq. (\ref{PayoffBob}), assuming $\theta _{\text{A,B}} = 0 $, we get:
\begin{eqnarray}
\overline{\$}_{\text{A}} \left( \theta _{\text{A}},\phi _{\text{A}} ,   \theta _{\text{B}} ,\phi _{\text{B}} \right) & = & 
\overline{\$}_{\text{A}} \left( 0,\phi _{\text{A}} ,  0,\phi _{\text{B}} \right) \ = \ 5\cos ^{2}\left( \phi _{\text{A}}+\phi _{\text{B}}\right) +3\sin ^{2} \left( \phi _{\text{A}}+\phi _{\text{B}}\right) \text{,} \quad  \quad  \\
\overline{\$}_{\text{B}}  \left( \theta _{\text{A}},\phi _{\text{A}} ,   \theta _{\text{B}} ,\phi _{\text{B}} \right) & = &
\overline{\$}_{\text{B}} \left( 0,\phi _{\text{A}} ,  0,\phi _{\text{B}} \right) \ = \ 3\cos ^{2}\left( \phi _{\text{A}}+\phi _{\text{B}}\right) +5\sin ^{2} \left( \phi _{\text{A}}+\phi _{\text{B}}\right) \text{.}  \quad   \quad 
\end{eqnarray}

Given that staying together favors both players {and is the most trivial case}, they should choose a quantum strategy operator that has the same $\theta$ and $\phi$ parameters {(the general case is demonstrated in appendix \ref{sec:TheoreticalAnalysis})}. From that, it is visible that the values of $\phi _{\text{A}}$ and  $\phi _{\text{B}}$  are the same and from the payoff equations they can be either $\pi/8$ or $3\pi/8$, yielding the possible quantum strategy operators {in the single qubit representation}:
\begin{eqnarray}
\widehat{\mathbf{U}}   \left(   0,    \frac{\pi }{8} \right) & = & \left[ 
\begin{array}{cc}
\exp \left[ i\frac{\pi }{8}\right]  & 0 \\ 
0 & \exp \left[ -i\frac{\pi }{8}\right] 
\end{array}%
\right]\text{,}  \label{Operator1stSolution} \\
\widehat{\mathbf{U}}  \left( 0 , \frac{3\pi }{8} \right) & = & \left[ 
\begin{array}{cc}
\exp \left[ i\frac{3\pi }{8}\right]  & 0 \\ 
0 & \exp \left[ -i\frac{3\pi }{8}\right] 
\end{array}%
\right]\text{,}   \label{Operator2ndSolution} 
\end{eqnarray}
{where the two player operator representation is $\widehat{\mathbf{U}}_{\text{A}} = \widehat{\mathbf{U}} \otimes \widehat{\mathbf{1}} $ and $\widehat{\mathbf{U}}_{\text{B}} =  \widehat{\mathbf{1}} \otimes \widehat{\mathbf{U}}$.  }
The strategy operators (\ref{Operator1stSolution}) and (\ref{Operator2ndSolution}) mean that Alice and Bob may choose the same entertainment option with payoffs that are equal for both. The payoff resulting from these strategies for each player is given by:
\begin{eqnarray}
\overline{\$}_{\text{A}} \left( \theta _{\text{A}} , \phi _{\text{A}} ,  \theta  _{\text{B}} , \phi _{\text{B}} \right) & = & 
\overline{\$}_{\text{A}} \left( 0,\frac{ \pi }{8} , 0,\frac{ \pi }{8} \right) \ = \ \frac{\alpha + \beta }{2} \ = \  \frac{5 + 3 }{2} \ = \ 4  \text{,}   \\
\overline{\$}_{\text{B}}  \left( \theta _{\text{A}} , \phi _{\text{A}} ,  \theta  _{\text{B}} , \phi _{\text{B}} \right) & = & 
\overline{\$}_{\text{B}} \left( 0,\frac{ \pi }{8} ,  0,\frac{ \pi }{8}  \right) \ = \ \frac{\beta + \alpha}{2} \ = \ \frac{3 + 5}{2} \ = \ 4  \text{.}  
\end{eqnarray}
Similar payoffs are achieved if the players use  $ \phi _{\text{A,B}} = \frac{3 \pi }{8} $.


With these results in hand, we can explore the Eisert-Wilkens-Lewenstein protocol to try and elucidate which actions to perform when confronted with any situation. Therefore, the protocol highlights all the available choices for each player such that it becomes visible which option minimizes losses or maximizes gains. From those possibilities,  the quantum game preserves the classical game with their respective payoff values, as we can see from data of Fig. \ref{fig:PayoffBattleOfSexes}, and  the most {trivial and} relevant moves are {summarized}  at the bi-matrix on Tab. \ref{tab:PayoffsBattleSexesNovo}. From the point past the Opera-Opera equilibria, we need a different interpretation to discuss and explain this strictly quantum regime of the game.

Fig. \ref{fig:PayoffBattleOfSexes}(a) and   Fig. \ref{fig:PayoffBattleOfSexes}(b), explicitly represent the bi-matrix of payoff, and from them, we can analyze some of the principal properties of game theory.

We distinguish three possible choices of quantum operator strategies that satisfy the pareto-optimal definition. This definition establishes that for a pair of strategies it is not possible to increase the payoff of one player without decreasing the payoff of another \cite{myerson1991Book,hofbauer1998Book,eisert1999}. {Mathematically, the rate of change of Alice's payoff with respect to any variation of one of their angular parameters has opposite signal to the rate of change of Bob's payoff. This statement is related with a geometrical interpretation of the following gradient: }
\begin{equation}
\left. \frac{\partial\overline{\$}_{\text{A}}}{\partial\xi _{\text{A}}}\right\vert
_{\left\{ \phi _{\text{A}}^{0},\theta _{\text{A}}^{0},\phi _{\text{B}%
}^{0},\theta _{\text{B}}^{0}\right\} }=-\left. \frac{\partial\overline{\$}_{\text{B}%
}}{\partial\xi _{\text{B}}} \right\vert _{\left\{\theta _{%
\text{A}}^{0}, \phi _{\text{A}}^{0},\theta _{\text{B}}^{0},\phi _{\text{B}}^{0}\right\} }\text{,}
\label{geometricalParetoOtimal}
\end{equation}%
{where $\left\{ \theta _{\text{A}}^{0},\phi _{\text{A}}^{0},\theta _{\text{B} }^{0},\phi _{\text{B}}^{0}\right\} $ is the bound established by the mathematical procedure at the end of appendix \ref{sec:TheoreticalAnalysis} and\ $\xi \in \left\{ \theta ,\phi \right\} $.}

Following the payoff curve from Fig. \ref{fig:PayoffBattleOfSexes}(c) {from a geometrical point of view and the 1$^{\circ}$, 2$^{\circ}$ and 4$^{\circ}$ items from the final part of appendix \ref{sec:TheoreticalAnalysis},}  we can see that the quantum strategy operators  $ \widehat{\mathbf{U}}_{\text{A}}\left( 0, \frac{ \pi }{8}  \right) $ and  $ \widehat{\mathbf{U}}_{\text{B}}\left( 0, \frac{ \pi }{8}  \right) $ satisfy the pareto-optimal definition, Eq. (\ref{geometricalParetoOtimal}), {so that computing the respective derivative operators the rates are found to be $  \frac{\partial\overline{\$}_{\text{A}}}{\partial\phi _{\text{A}}} = - \alpha+\beta $ and $  \frac{\partial\overline{\$}_{\text{B}}}{\partial\phi _{\text{B}}} =  \alpha-\beta $ at the bound parameter values  $\left\{ \theta _{\text{A}}^{0}=0,\phi _{\text{A}}^{0} = \frac{\pi}{8},\theta _{\text{B} }^{0}=0,\phi _{\text{B}}^{0} = \frac{\pi}{8}\right\} $}. The same procedure is applied for the quantum strategy operator $ \widehat{\mathbf{U}}_{\text{A}}\left( 0, \frac{3 \pi }{8}  \right) $ with  $ \widehat{\mathbf{U}}_{\text{B}}\left( 0, \frac{3 \pi }{8}  \right) $, { in which the rates are found  $  \frac{\partial\overline{\$}_{\text{A}}}{\partial\phi _{\text{A}}} =  \alpha-\beta $ and $  \frac{\partial\overline{\$}_{\text{B}}}{\partial\phi _{\text{B}}} = -\alpha+\beta $ at the bound parameter values  $\left\{ \theta _{\text{A}}^{0}=0,\phi _{\text{A}}^{0} = \frac{3\pi}{8},\theta _{\text{B} }^{0}=0,\phi _{\text{B}}^{0} = \frac{3\pi}{8}\right\} $}. In the case of  the quantum strategy operator $ \widehat{\mathbf{U}}_{\text{A}}\left(\frac{ \pi }{2} , 0 \right) $ with $ \widehat{\mathbf{U}}_{\text{B}}\left(\frac{ \pi }{2} , 0 \right) $ { the rates are  $  \frac{\partial\overline{\$}_{\text{A}}}{\partial\theta _{\text{A}}} =  \frac{-\alpha+\beta}{4} $ and $  \frac{\partial\overline{\$}_{\text{B}}}{\partial\theta _{\text{B}}} = \frac{\alpha-\beta}{4} $ at the bound parameter values  $\left\{ \theta _{\text{A}}^{0}= \frac{\pi}{2},\phi _{\text{A}}^{0} =0,\theta _{\text{B} }^{0}= \frac{\pi}{2},\phi _{\text{B}}^{0} =0\right\} $}.

One interesting point about these three pairs of quantum strategy operators is that the payoffs for both players are the same. However, the payoff values $ \overline{\$} _ {\text{A}}\left( 0, \frac{ \pi }{8} , 0 ,   \frac{ \pi }{8} \right)  =  \overline{\$} _ { \text{B} } \left( 0, \frac{ \pi }{8} , 0 ,   \frac{ \pi }{8} \right) =  \frac{ \alpha + \beta }{ 2 }$ and  $ \overline{\$} _ {\text{A}}\left( 0, \frac{3\pi }{8} , 0 ,   \frac{3 \pi }{8} \right)  =  \overline{\$} _ { \text{B} } \left( 0, \frac{3\pi }{8} , 0 ,   \frac{3\pi }{8} \right) =  \frac{ \alpha + \beta }{ 2 }$  while  $ \overline{\$} _ { \text{A} } \left( \frac{\pi }{2} , 0 ,   \frac{\pi }{2} , 0 \right)  =  \overline{\$} _ { \text{B} } \left( \frac{\pi }{2} , 0 ,   \frac{\pi }{2} , 0 \right) =  \frac{\alpha + \beta + 2 \gamma}{4}$ such that   $  \frac{ \alpha + \beta }{ 2 } > \frac{\alpha + \beta + 2 \gamma}{4}$. The latter set of angular parameters $ \lbrace \theta _{\text{A}}^{0}=\frac{\pi }{2} , \phi _{\text{A}}^{0}=0 ,\theta _{\text{B}}^{0}= \frac{\pi }{2} , \phi_{\text{B}}^{0}=0 \rbrace $ is not so interesting because the degree of satisfaction of both players is considerably smaller than the degree of satisfaction of the former.

Those set of angular parameters configure pairs of quantum strategy operators that players can apply to achieve the same degree of happiness, without restricting their moves as happens in the original version of Marinatto-Weber protocol \cite{marinatto2000} and in other studies \cite{pappa2015,brunner2013}.

{There is another interesting set of angular parameter values established by equations  $ \tan \frac{\theta _{\text{A}}}{2}  =  \cot \frac{\theta _{\text{B}}}{2}$  and $ \phi_{\text{A}} + \phi_{\text{B}} =  \frac{\pi }{2} $ (see item  3$^{\circ}$ at the end of appendix \ref{sec:TheoreticalAnalysis}). The player's payoffs are analysed using the linear equation, making the player's payoffs dependent only on  $ \theta _{\text{A}} $ and $ \theta _{\text{B}} $. Next, computing the maximum or minimum of any function, it  shows that Alice's and Bob's payoff achieve their  maximum at  $\theta _{\text{A}} = \theta _{\text{B}}  $. This equality is used to solve the equation  $ \tan \frac{\theta _{\text{A}}}{2}  =  \cot \frac{\theta _{\text{B}}}{2}$, where the angular parameters satisfy $\theta _{\text{A}} = \theta _{\text{B}} = \frac{\pi}{2}  $ and represent a maximum. This set of angular parameters is used to verify if the player's payoffs satisfy  the definition of Nash equilibrium (see Eq. (\ref{NashEquiAlice}) and Eq. (\ref{NashEquiBob})), so that we  found many Nash equilibria given by  $ \phi_{\text{A}} + \phi_{\text{B}} =  \frac{\pi }{2} $ and  $\theta _{\text{A}} = \theta _{\text{B}} = \frac{\pi}{2}  $. A graphical representation of them is made on Fig.  \ref{fig:ManyAngularParametersPayoffs}(d,e,f) where the maximum of the surface achieves the definition, and the trivial case  $ \phi_{\text{A}} = \phi_{\text{B}} =  \frac{\pi }{4} $ and  $\theta _{\text{A}} = \theta _{\text{B}} = \frac{\pi}{2}  $, is represented by a black circle in Fig.  \ref{fig:ManyAngularParametersPayoffs}(f).  Similar results were summarized on table of Sec. IIB of Ref. \cite{du2001ppNovembro}.}

Lastly, there is a common misunderstanding related to Table \ref{tab:PayoffsBattleSexesNovo}, when using it to define a new classical game \cite{van-Enk2002}.  If one desires to map a classical game from a quantum game, it is necessary to use the data from Fig.  \ref{fig:PayoffBattleOfSexes} or Eq. (\ref{PayoffAlice}) and Eq. (\ref{PayoffBob}). The advantage of the quantum game arises at this point. We understand that using Table  \ref{tab:PayoffsBattleSexesNovo} to map a classical game is a tricky procedure because it is easy to map a classical game from a quantum game if one knows the solutions, but the mapping becomes more laborious if one does not know the solution. In this  work, we use  Table \ref{tab:PayoffsBattleSexesNovo} only to show numerical values of payoffs related to Fig. \ref{fig:PayoffBattleOfSexes} for the most important quantum strategy operators to elucidate the solution of the game. {Another interesting point in this discussion is related to  the effect of the entanglement parameter. Using similar procedures and considering trivial angular parameters values, we find that $\lambda = \frac{\pi}{4}$ is the minimum value at which the present discussion makes sense. Below this value, the pareto-otimal solutions are lost and the advantages of entanglement drastically diminish.} \textcolor{magenta}{}

\section{Conclusions}\label{sec:Conclusions}

In summary, we have explored the Eisert-Wilkens-Lewenstein protocol to analyze an asymmetric game, the battle of the sexes. It preserves the classical regime and also gives insights at the quantum regime. Two pareto-optimal solutions that provide the same degree of satisfaction for both players arise. The degree of satisfaction for both players is compatible with that predicted from Marinatto-Weber protocol. Quantum states that represent the solution for the battle of the sexes are not solutions to the prisoner's dilemma game. Finally, the Eisert-Wilkens-Lewenstein protocol may be successfully implemented even in the case of asymmetric games at the quantum description, not being restricted to the symmetric cases.

\begin{acknowledgements}
The authors  acknowledge  the National Institute of Science and Technology for Quantum Information (INCT-QI). A.C.S.L. acknowledges  financial support from  CNPq (142118/2018-4). E.L.O.  acknowledges financial support from  CNPq (140215/2015-8).  T.J.B. acknowledges financial support from CNPq (308488/2014-8) and FAPESP (2012/02208-5). R.A. acknowledges  financial support from CNPq (309023/2014-9, 459134/2014-0). This study was financed in part by the Coordena\c{c}\~{a}o de Aperfei\c{c}oamento de Pessoal de N\'{i}vel Superior - Brasil (CAPES) - Finance Code 001.
\end{acknowledgements}

\begin{appendices}

\section{Theoretical analysis}\label{sec:TheoreticalAnalysis}

{Extended details are presented in this appendix to highlight the importance of the theoretical and experimental data  contained in Fig. \ref{fig:PayoffBattleOfSexes}(a,b,c). Also, we give some details about the  mathematical procedures from which we draw our conclusions.}
 
{The most relevant aspect of quantum circuits is their ability to provide analytical solutions. For this reason, we test  the   Eisert-Wilkens-Lewenstein protocol \cite{eisert1999} in the context of asymmetrical games. As it was introduced in the main text, the protocol delivers mathematical expressions of probabilities at different degrees of entanglement $\left( \lambda \right) $,  denoted by $ P_{s_{\text{A}},s_{\text{B}}}=\left\vert \left\langle s_{\text{A}},s_{\text{B }}|\psi _{f}\right\rangle \right\vert ^{2}$, where  $ s_{\text{A,B}} \in \left\lbrace O , T \right\rbrace   $, such that the probabilities are explicitly expressed } 
\begin{eqnarray}
P_{OO} & = & \left( \cos ^{2}\left( \phi _{%
\text{A}}+\phi _{\text{B}}\right) +\sin ^{2}\left( \phi _{\text{A}}+\phi _{  \text{B}}\right) \cos ^{2}\lambda \right) \cos ^{2}\frac{\theta _{\text{A}}}{ 2}\cos ^{2}\frac{\theta _{\text{B}}}{2}  \text{,} \label{POOWithEntangledParameters}    \\
P_{OT} & = & \left( \cos ^{2}\phi _{\text{A}}+\sin ^{2}\phi _{\text{A} } \cos ^{2}\lambda \right) \cos ^{2}\frac{\theta _{\text{A}}}{2}\sin ^{2} \frac{\theta _{\text{B}}}{2}-\frac{\cos \phi _{\text{A}}\sin \phi _{\text{B}  } \sin \theta _{\text{A}}\sin \theta _{\text{B}}\sin \lambda }{2} +  \left(  \sin  \phi _{\text{B}}\sin  \frac{\theta _{\text{A}}}{2}\cos  \frac{ \theta _{\text{B}}}{2}\sin\lambda  \right)   ^{2} \text{,}  \qquad . \label{POTWithEntangledParameters}    \\
P_{TO} & = & \left( \cos ^{2}\phi _{\text{B}}+\sin ^{2}\phi _{\text{B} } \cos ^{2}\lambda \right) \sin ^{2}\frac{\theta _{\text{A}}}{2}\cos ^{2} \frac{\theta _{\text{B}}}{2}-\frac{\cos \phi _{\text{B}}\sin \phi _{\text{A} } \sin \theta _{\text{A}}\sin \theta _{\text{B}}\sin \lambda }{2} + \left(  \sin  \phi _{\text{A}}\cos  \frac{\theta _{\text{A}}}{2}\sin  \frac{ \theta _{\text{B}}}{2}\sin\lambda  \right) ^{2} \text{,} \qquad  . \label{PTOWithEntangledParameters}   \\
P_{TT} & = & \left( \sin \frac{\theta _{\text{A}}}{2}\sin \frac{\theta _{ \text{B} } }{2} + \sin \left( \phi _{\text{A}}+\phi _{B}\right) \cos \frac{ \theta _{ \text{A} } }{2} \cos \frac{\theta _{\text{B}}}{2}\sin \lambda \right) ^{2}  \text{.} \label{PTTWithEntangledParameters}
\end{eqnarray}
{We apply Eq. (\ref{POOWithEntangledParameters}-\ref{PTTWithEntangledParameters}) to evaluate Alice's and Bob's payoffs established in Eq. (\ref{PayoffAlice}) and Eq. (\ref{PayoffBob}), respectively. Both expressions depend on eight parameters $\phi _{\text{A}}$, $\phi _{\text{B}}$, $\theta _{\text{A}}$, $\theta _{\text{B}}$, $\lambda$, $\alpha$, $\beta$, and $\gamma$. In this paper, we analyse the player's payoffs at the highest degree of entanglement with $\lambda=\frac{\pi}{2}$, allowing  Eq. (\ref{PayoffAlice}) and Eq. (\ref{PayoffBob}) are rewritten}
\begin{eqnarray*}
\overline{\$}_{\text{A}} &=&\alpha \left(  \cos ^{2}\left( \phi _{ \text{ A } } + \phi _{\text{B}}\right)  \cos ^{2}\frac{\theta _{\text{A}}}{ 2 } \cos ^{2}\frac{\theta _{\text{B}}}{2}\right) \\
&&+\gamma \left(   \cos ^{2}\phi _{\text{A}}  \cos ^{2}\frac{ \theta _{\text{A}}}{2}\sin ^{2}\frac{\theta _{\text{B}}}{2}-\frac{\cos \phi _{ \text{A} } \sin \phi _{\text{B}} \sin \theta _{\text{A}} \sin \theta _{\text{B}  } } {2}+\sin ^{2}\phi _{\text{B}}\sin ^{2}\frac{\theta _{\text{A}}}{2}\cos ^{2} \frac{\theta _{\text{B}}}{2}\right) \\
&&+\gamma \left(   \cos ^{2}\phi _{\text{B}}  \sin ^{2}\frac{ \theta _{\text{A}}}{2}\cos ^{2}\frac{\theta _{\text{B}}}{2}-\frac{\cos \phi _ { \text{B} }\sin \phi _{\text{A}}\sin \theta _{\text{A}}\sin \theta _{\text{B} } } {2}+\sin ^{2}\phi _{\text{A}}\cos ^{2}\frac{\theta _{\text{A}}}{2}\sin ^{2} \frac{\theta _{\text{B}}}{2}\right) \\
&&+\beta   \left( \sin \frac{\theta _{\text{A}}}{2}\sin \frac{\theta _{ \text{B}}}{2}+\sin \left( \phi _{\text{A}}+\phi _{\text{B}}\right) \cos \frac{ \theta _{\text{A}}}{2}\cos \frac{\theta _{\text{B}}}{2}\right) ^ { 2 }    \text{,}
\end{eqnarray*}
\begin{eqnarray*}
\overline{\$}_{\text{B}} &=&\beta \left(   \cos ^{2}\left( \phi _{\text{%
A}}+\phi _{\text{B}}\right)   \cos ^{2}\frac{\theta _{\text{A}}}{2}%
\cos ^{2}\frac{\theta _{\text{B}}}{2}\right) \\
& &+\gamma \left(   \cos ^{2}\phi _{\text{A}}  \cos ^{2}\frac{%
\theta _{\text{A}}}{2}\sin ^{2}\frac{\theta _{\text{B}}}{2}-\frac{\cos \phi
_{\text{A}}\sin \phi _{\text{B}}\sin \theta _{\text{A}}\sin \theta _{\text{B}%
}}{2}+\sin ^{2}\phi _{\text{B}}\sin ^{2}\frac{\theta _{\text{A}}}{2}\cos ^{2}%
\frac{\theta _{\text{B}}}{2}\right) \\
& &+\gamma \left(   \cos ^{2}\phi _{\text{B}}  \sin ^{2}\frac{%
\theta _{\text{A}}}{2}\cos ^{2}\frac{\theta _{\text{B}}}{2}-\frac{\cos \phi
_{\text{B}}\sin \phi _{\text{A}}\sin \theta _{\text{A}}\sin \theta _{\text{B}%
}}{2}+\sin ^{2}\phi _{\text{A}}\cos ^{2}\frac{\theta _{\text{A}}}{2}\sin ^{2}%
\frac{\theta _{\text{B}}}{2}\right) \\
& &+\alpha   \left( \sin \frac{\theta _{\text{A}}}{2}\sin \frac{\theta _{%
\text{B}}}{2}+\sin \left( \phi _{\text{A}}+\phi _{\text{B}}\right) \cos 
\frac{\theta _{\text{A}}}{2}\cos \frac{\theta _{\text{B}}}{2}\right)
^{2}  \text{,}
\end{eqnarray*}
{so that, if  $\gamma = 0$  these mathematical expressions of both players payoffs match analogous expressions as in Eq. (3.2) of Ref. \cite{alonso-sanz2012}.} 

{With these expressions in hand,  we perform the first stage of the mathematical procedure to find Pareto-optimal solutions, which is setting $\overline{\$}_{\text{A}}=\overline{\$}_{\text{B}}$, or equivalently $\overline{\$}_{\text{A}}-\overline{\$}_{\text{B}}=0$. After an algebraic manipulation we obtain }
\begin{equation}
\left(   \cos \left( \phi _{\text{A}}+\phi _{\text{B}}  \right)
\cos \frac{\theta _{\text{A}}}{2}\cos \frac{\theta _{\text{B}}}{2}\right)
^{2}-\left( \sin \frac{\theta _{\text{A}}}{2}\sin \frac{\theta _{\text{B}}}{2%
}+\sin \left( \phi _{\text{A}}+\phi _{\text{B}}\right) \cos \frac{\theta _{%
\text{A}}}{2}\cos \frac{\theta _{\text{B}}}{2}\right) ^{2}=0\text{,}
\label{GeneralExpression}
\end{equation}
{as we see, the main advantage of the algebraic manipulation is to reduce the parameter dependence to only  $\phi _{\text{A}}$, $\phi _{\text{B}}$, $\theta _{\text{A}}$, $\theta _{\text{B}}$. Next we separate the parameter dependence by generating a quadratic polynomial in $ \tan \frac{\theta _{\text{A}}}{2}\tan \frac{\theta _{\text{B}}}{2}$ with the help of trigonometrical identities and maintaining  $ \phi _{\text{A}}$ and $\phi _{\text{B}}$ as coefficients of the quadratic polynomial}
\begin{equation}
\left( \tan \frac{\theta _{\text{A}}}{2}\tan \frac{\theta _{\text{B}}}{2}\right) ^{2} 
+2\sin \left( \phi _{\text{A}}+\phi _{\text{B}} \right) \tan \frac{\theta _{\text{A}}}{2}\tan \frac{\theta _{\text{B}}}{2} 
-\left( \cos ^{2}\left( \phi _{\text{A}}+\phi _{\text{B}}\right) -\sin ^{2}\left( \phi _{\text{A}}+\phi _{\text{B}}\right) \right) =  0\text{,}
\end{equation}
{and solving the quadratic equation we find the roots: }
\begin{eqnarray}
\tan \frac{\theta _{\text{A}}}{2}\tan \frac{\theta _{\text{B}}}{2} &=&-\sin
\left( \phi _{\text{A}}+\phi _{\text{B}}\right) +\cos \left( \phi _{\text{A}%
}+\phi _{\text{B}}\right) \text{,} \label{BhaskarasSolutionMas} \\
\tan \frac{\theta _{\text{A}}}{2}\tan \frac{\theta _{\text{B}}}{2} &=&-\sin
\left( \phi _{\text{A}}+\phi _{\text{B}}\right) -\cos \left( \phi _{\text{A}%
}+\phi _{\text{B}}\right) \text{,} \label{BhaskarasSolutionMenos}
\end{eqnarray}
{ Eq. (\ref{BhaskarasSolutionMas}) and Eq. (\ref{BhaskarasSolutionMenos})   represent two independent mathematical expressions. To correlate them, we multiply  and take the square root of each side, so that we have:}
\begin{equation}
\tan \frac{\theta _{\text{A}}}{2}\tan \frac{\theta _{\text{B}}}{2}  = \sqrt{ 1-2\cos ^{2}\left( \phi _{\text{A}}+\phi _{\text{B}}\right) }  = \sqrt{ -\cos 2\left( \phi _{\text{A}}+\phi _{\text{B}}\right) } = \sqrt{\Phi  \left( \phi _{\text{A}}+\phi _{\text{B}}\right) } \text{.} \label{BhaskarasSolutionCorrelated}
\end{equation}
{The two key characteristics of Eq. (\ref{BhaskarasSolutionCorrelated}) are that ($i$) two pairs of angular parameters are at opposite sides of the expression, and the analysis can be made separately; ($ii$) taking the square root of both sides makes the valid domain of $\phi$ values easily visible since the right side must remain a real number ($\mathbb{R}$).  In this sense,  we first analyze the term with dependence on $ \phi _{\text{A}}+\phi _{\text{B}} $ and next the term  with dependence on $\theta _{\text{A}}$ and $\theta _{\text{B}}$.}

{The  $\phi _{\text{A}}$ and $\phi _{\text{B}}$ angular parameter dependence on the right side of  Eq. (\ref{BhaskarasSolutionCorrelated}) can be analysed by estimating their roots and their maximum and minimum values}
\begin{equation}
 -\cos 2\left( \phi _{\text{A}}+\phi _{\text{B}}\right) = \Phi  \left( \phi _{\text{A}}+\phi _{\text{B}}\right) \geq 0 \text{.} \label{BhaskarasSolutionFactors}
\end{equation}
{where the linear equation satisfies the following bounds}
\begin{equation}
\phi _{\text{A}}+\phi _{\text{B}}=\left\{ 
\begin{array}{ccc}
\frac{\pi }{4},\frac{3\pi }{4} &  & \text{root values of }  \Phi   \\ 
\frac{\pi }{2} &  & \text{maximum value of }  \Phi   \\ 
0,\pi  &  & \text{minimum values of }  \Phi    %
\end{array}%
\right.   \label{valores}
\end{equation}
{We can establish the parameter dependence with the following function for any value of  $ \left( \phi _{\text{A}}+\phi _{\text{B}}\right) \in  \left[ 0 , \pi  \right]  $:}
\begin{equation}
\tan \frac{\theta _{\text{A}}}{2}\tan \frac{\theta _{\text{B}}}{2}=\left\{ 
\begin{array}{ccc}
-\sin \left( \phi _{\text{A}}+\phi _{\text{B}}\right) +\cos \left( \phi _{%
\text{A}}+\phi _{\text{B}}\right)  & \text{,} & \phi _{\text{A}}+\phi _{%
\text{B}}\in \left[ 0,\frac{\pi }{4}\right] \cup \left[ \frac{3\pi }{4},\pi %
\right]  \\ 
&  &  \\ 
\sqrt{-\cos 2\left( \phi _{\text{A}}+\phi _{\text{B}}\right) } & \text{,} & 
\phi _{\text{A}}+\phi _{\text{B}}\in \left[ \frac{\pi }{4},\frac{3\pi }{4}%
\right] 
\end{array}%
\right.   \label{SolutionsParametersA}
\end{equation}

\begin{SCfigure}
\includegraphics[width=0.50\textwidth]{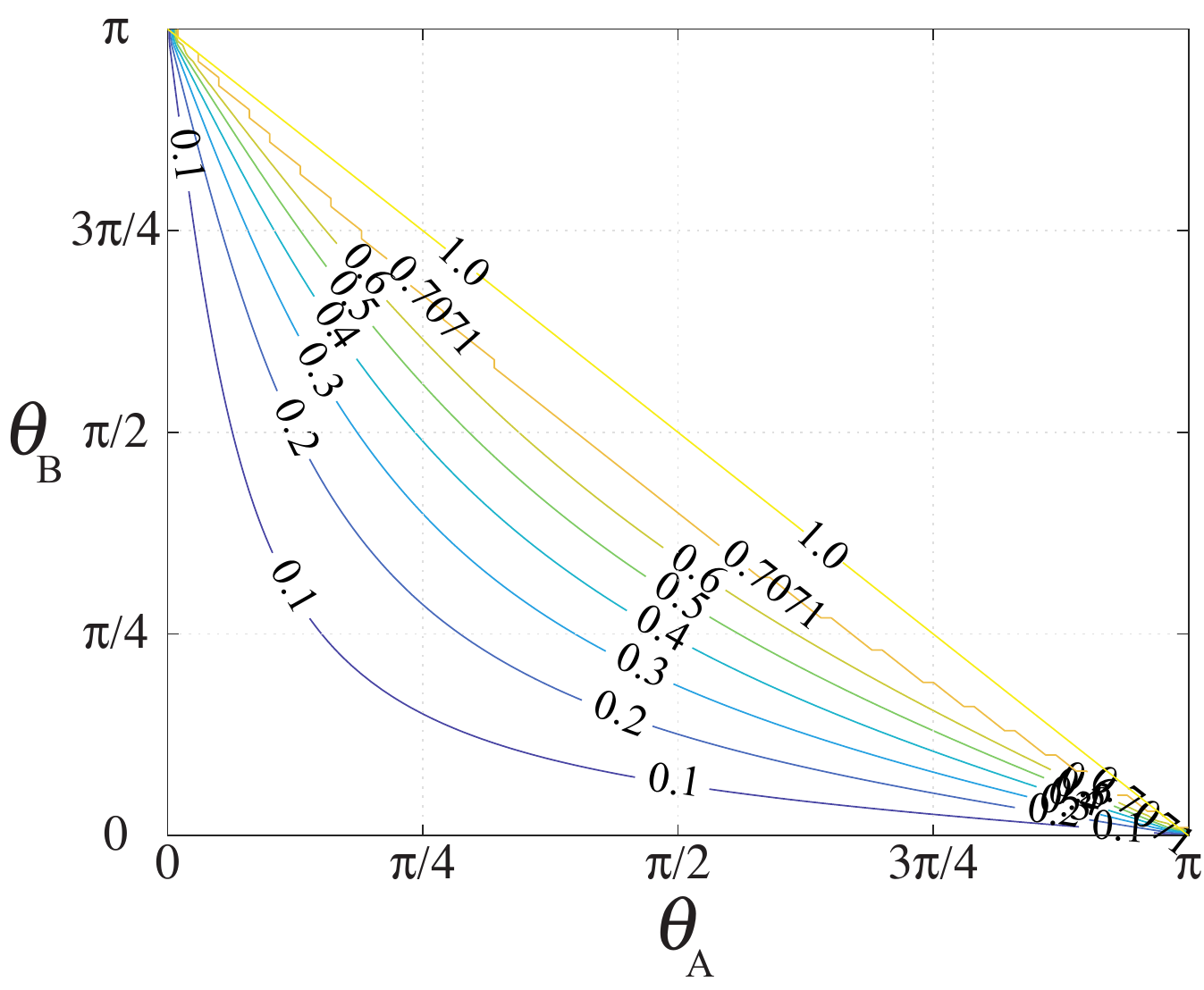}
\caption{(Color online) We sketch the left side of Eq.   (\ref{BhaskarasSolutionCorrelated}) as a contour plot, and we use eight   bounding values  of   $ \sqrt{\Phi  \left( \phi _{\text{A}} + \phi _{\text{B}}\right)} = 0.1, \ 0.2, \ \ldots \ , \ \sqrt{\frac{1}{2}}, 1 $, which the last one corresponds to the upper bound of real values.} \label{fig:FunctionTangent}
\end{SCfigure}

{The $  \theta _{\text{A}}$  and $\theta _{\text{B}}  $ angular parameter dependence on the left side of the Eq.  (\ref{BhaskarasSolutionCorrelated}) can be analysed using the bounds established by the procedure on the right side of Eq. (\ref{BhaskarasSolutionCorrelated}). We sketched them in Fig. \ref{fig:FunctionTangent} as  a  contour plot evaluating Eq. (\ref{BhaskarasSolutionCorrelated}) under eight  values for   $\sqrt{ \Phi  \left( \phi _{ \text{A} } + \phi _{ \text{B} } \right) }  =   0.1, \ 0.2, \ \ldots \ , \ \sqrt{\frac{1}{2}}, 1  $. The main feature of the equation in  Eq. (\ref{BhaskarasSolutionCorrelated}) is that  as  $\theta _{\text{A}} $ $  \left( \theta _{ \text{B} } \right)  $    increases   $ \theta _{\text{B}}  $ $  \left( \theta _{ \text{A} } \right)  $ decreases, enabling many different solutions. We will focus only on the special values related with   $\sqrt{ \Phi  \left( \phi _{ \text{A} } + \phi _{ \text{B} } \right) }$. For the lower bound of $\sqrt{ \Phi  \left( \phi _{ \text{A} } + \phi _{ \text{B} } \right) } = 0$, the   $  \theta _{\text{A}}$  and $\theta _{\text{B}}  $ angular parameters are null. For the upper bound of $\sqrt{ \Phi  \left( \phi _{ \text{A} } + \phi _{ \text{B} } \right) } = 1$, the    $  \theta _{\text{A}}$  and $\theta _{\text{B}}  $ angular parameters are related by the trigonometric equation $ \tan \frac{\theta _{\text{A}}}{2}\tan \frac{\theta _{\text{B}}}{2}  = 1$ or   $ \tan \frac{\theta _{\text{A}}}{2}  =  \cot \frac{\theta _{\text{B}}}{2}$, and this one is sketched in Fig. \ref{fig:FunctionTangent}. In the case of  $\sqrt{\Phi  \left( \phi _{\text{A}} + \phi _{\text{B}}\right)}  = i  $, the analysis is made on  Eq. (\ref{BhaskarasSolutionMas}) and Eq. (\ref{BhaskarasSolutionMenos}) which means weak correlated parameter values between  $ \left\{ \phi _{\text{A}} , \phi _{\text{B}}   \right\}  $ and  $ \left\{ \theta _{\text{A}} , \theta _{\text{B}}   \right\}  $. From this procedure there are two equations $ \tan \frac{\theta _{\text{A}}}{2}\tan \frac{\theta _{\text{B}}}{2}  = \pm 1$. } 

\begin{figure}[!ht]
\begin{center}
$
\begin{array}{c@{\hspace{0.10in}}c@{\hspace{0.10in}}c}	\\ [-0.53cm] 
\epsfxsize=1.50in \epsffile{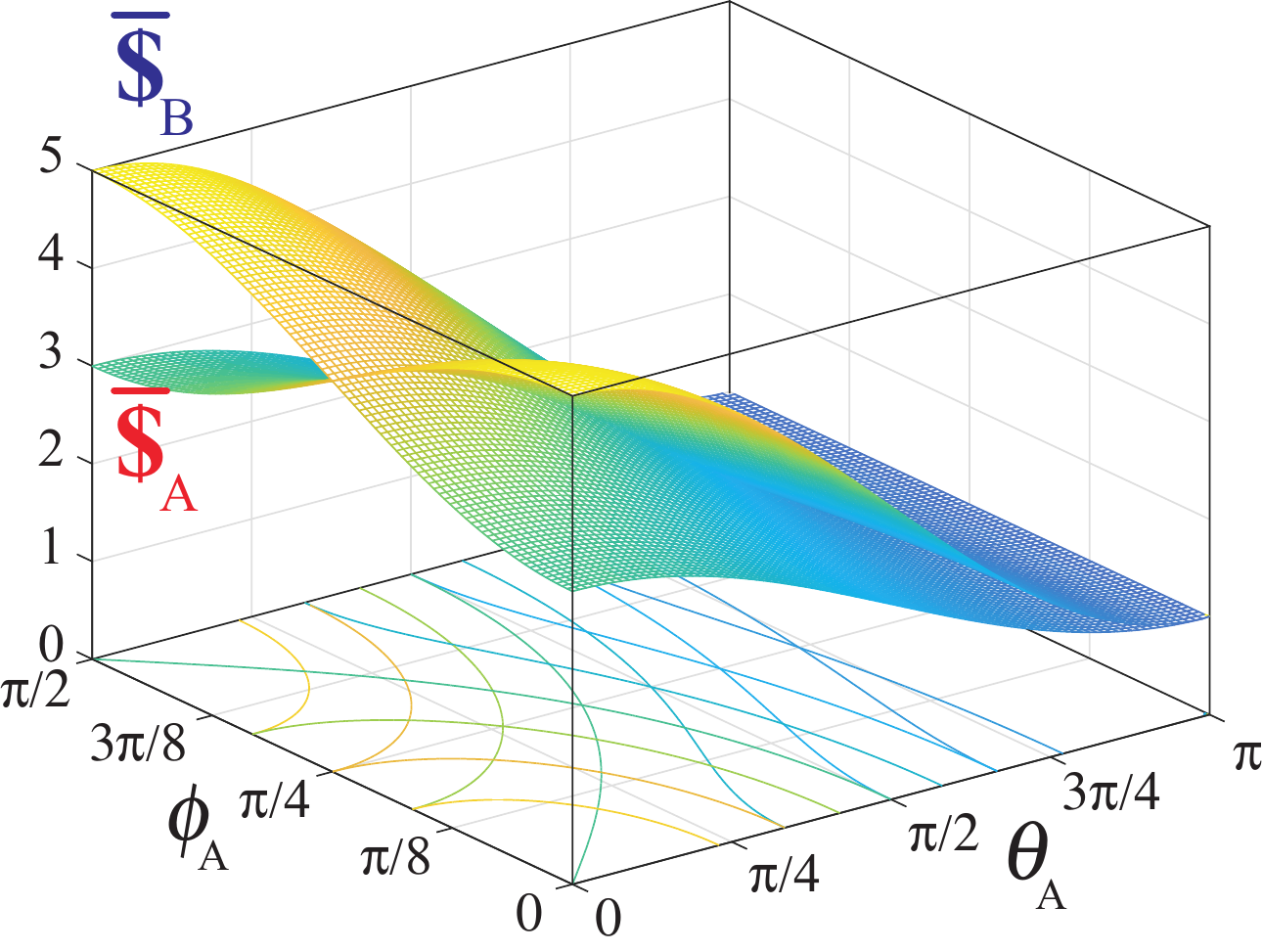}	& 
\epsfxsize=1.50in \epsffile{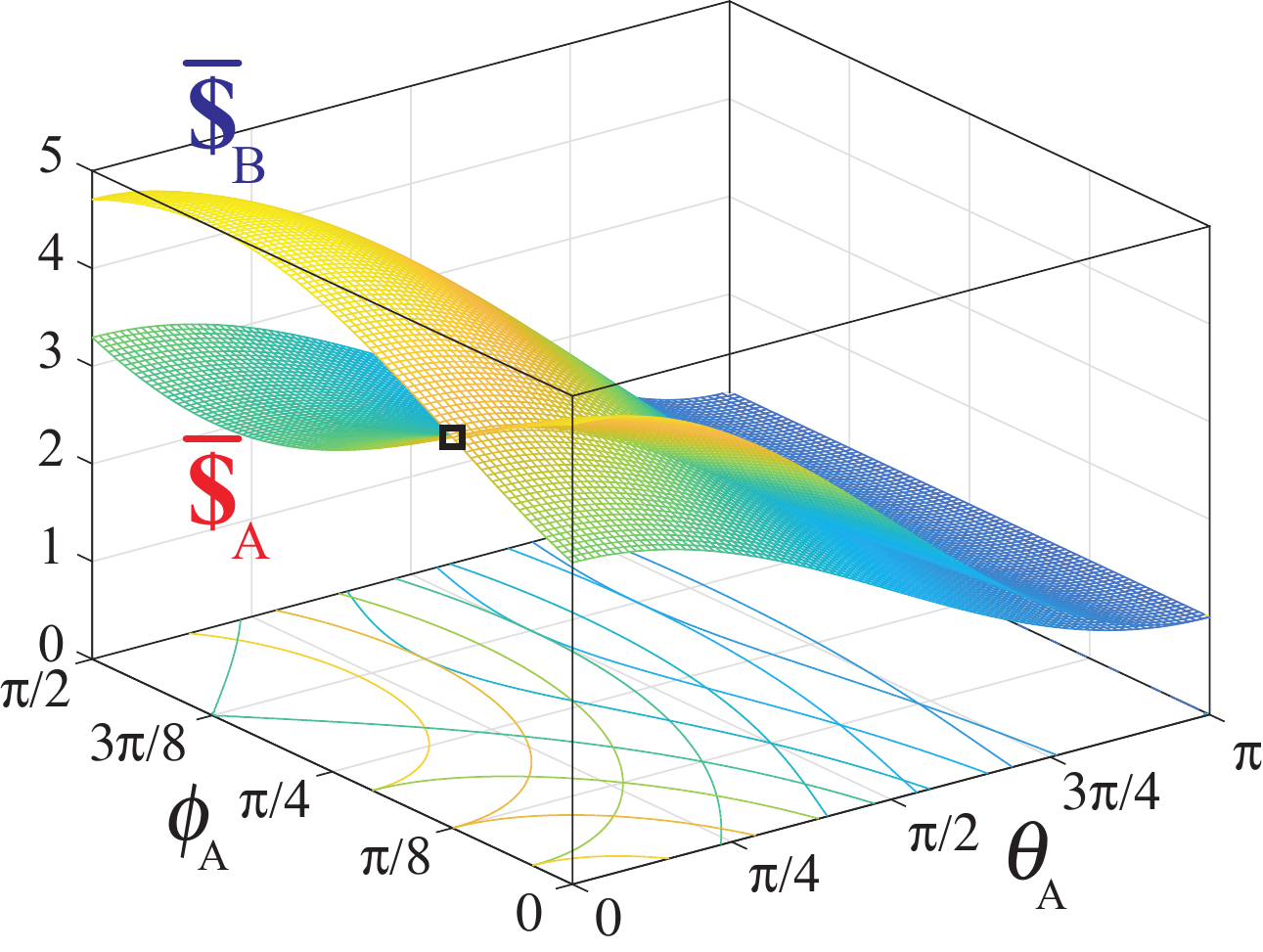}	& 
\epsfxsize=1.50in \epsffile{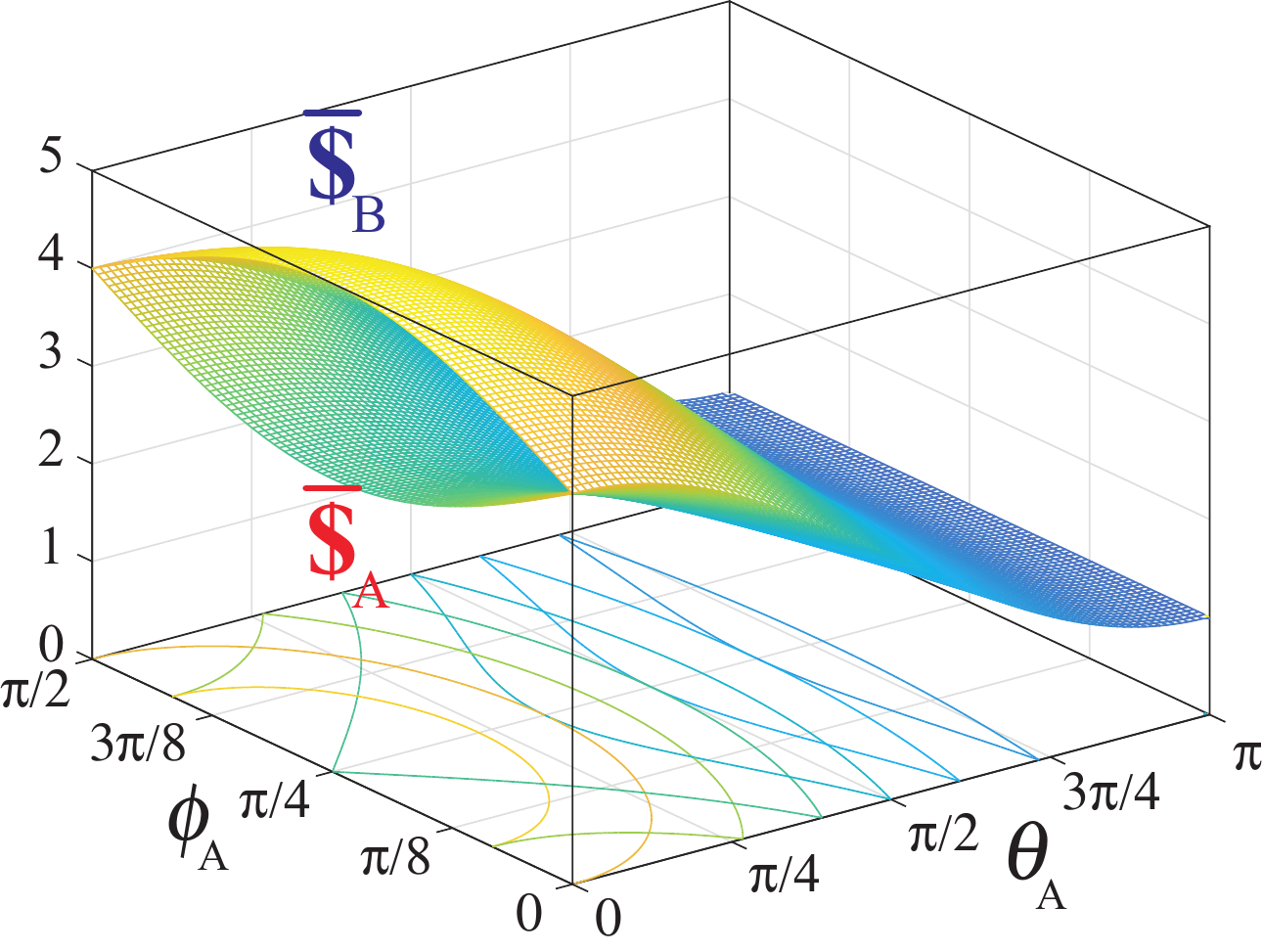}	\\ [0.00cm] 
\mbox{(a) $\phi_{\text{B}} =   0   \quad \theta_{\text{B}} = 0 $ }   &   \mbox{(b) $\phi_{\text{B}} = \pi/8 \quad \theta_{\text{B}} = 0 $ }   &   \mbox{(c) $\phi_{\text{B}} = \pi/4 \quad \theta_{\text{B}} = 0 $ }	\\ [0.00cm]
\epsfxsize=1.50in \epsffile{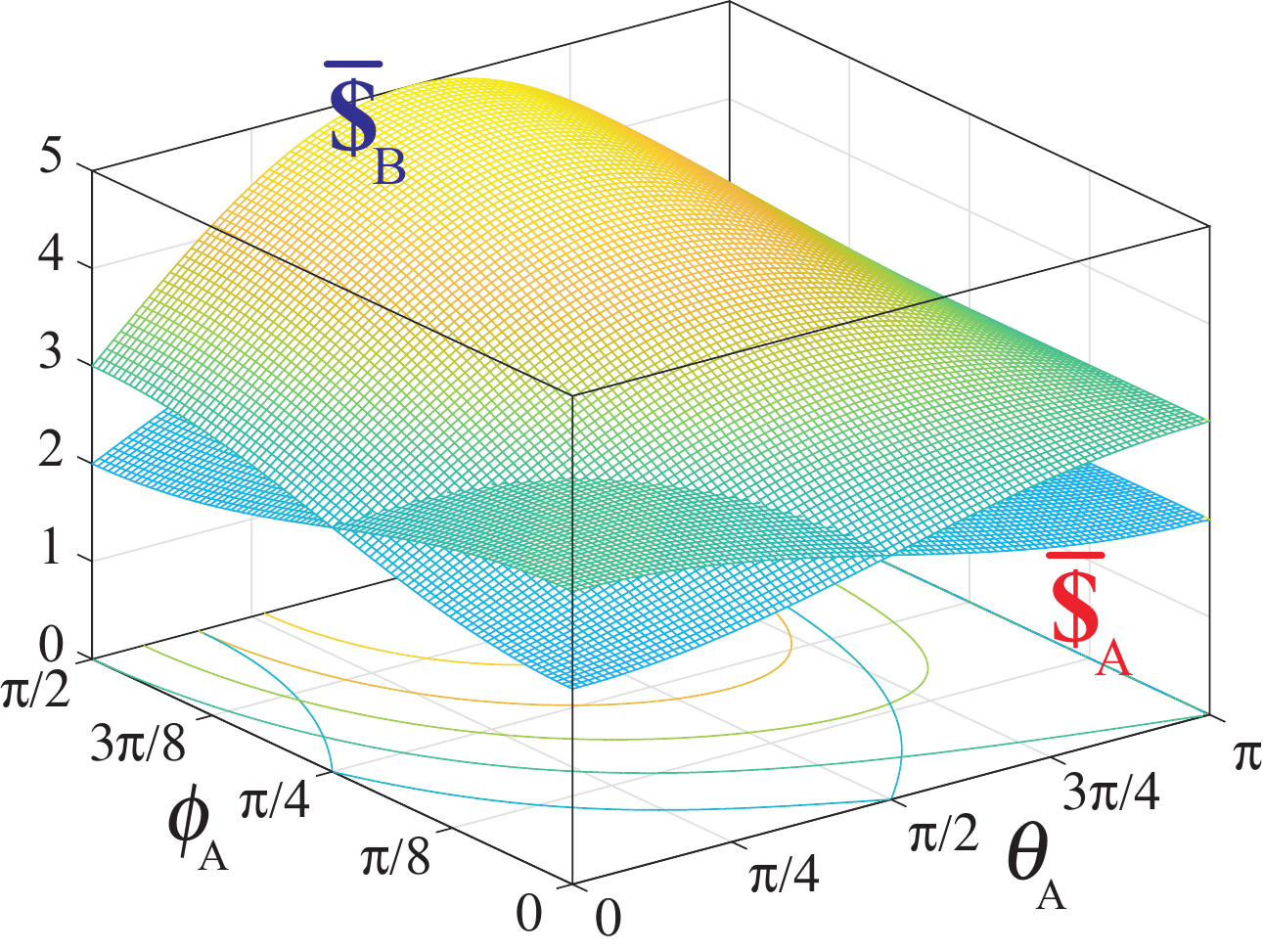}	& 
\epsfxsize=1.50in \epsffile{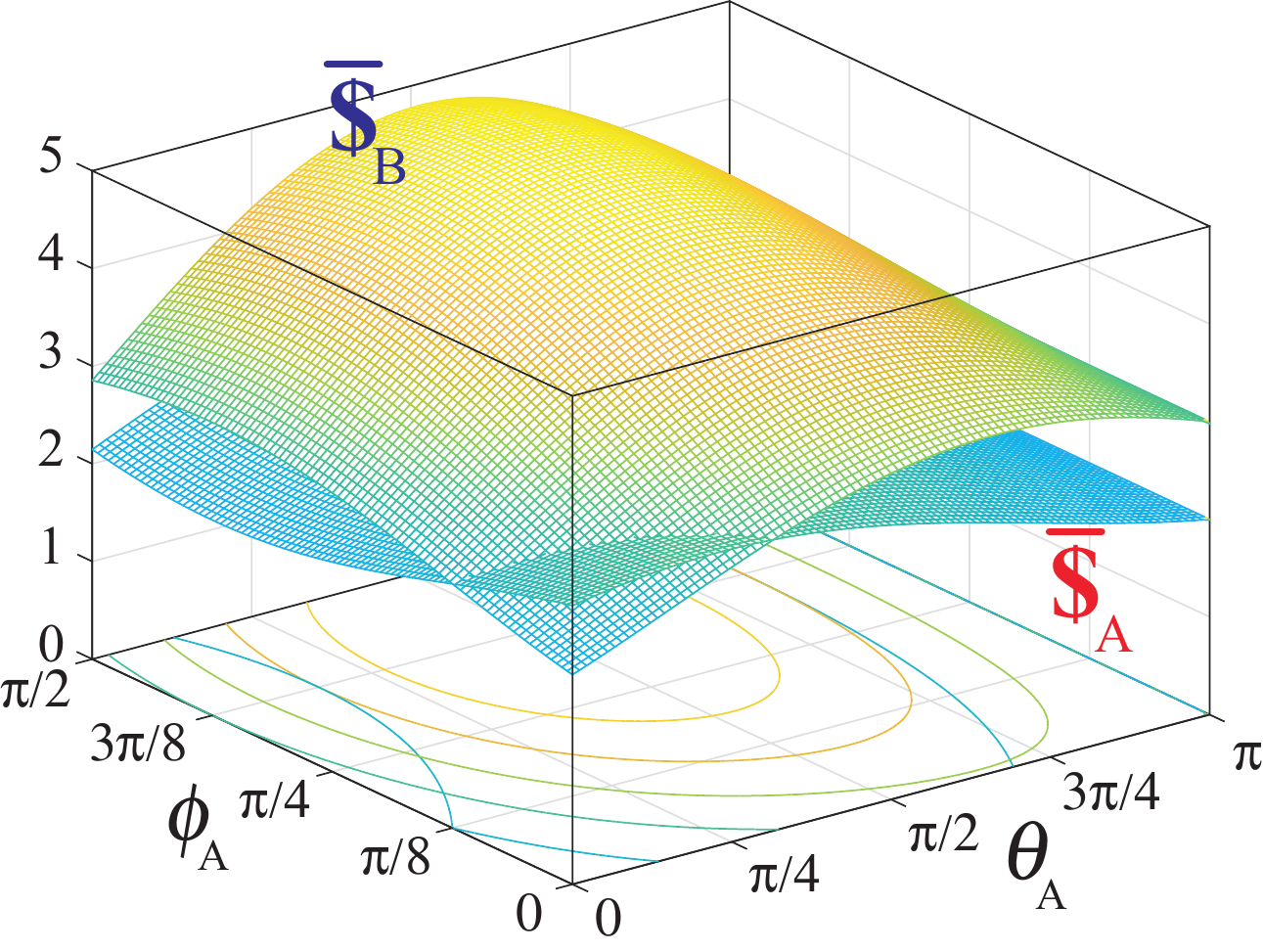}	& 
\epsfxsize=1.50in \epsffile{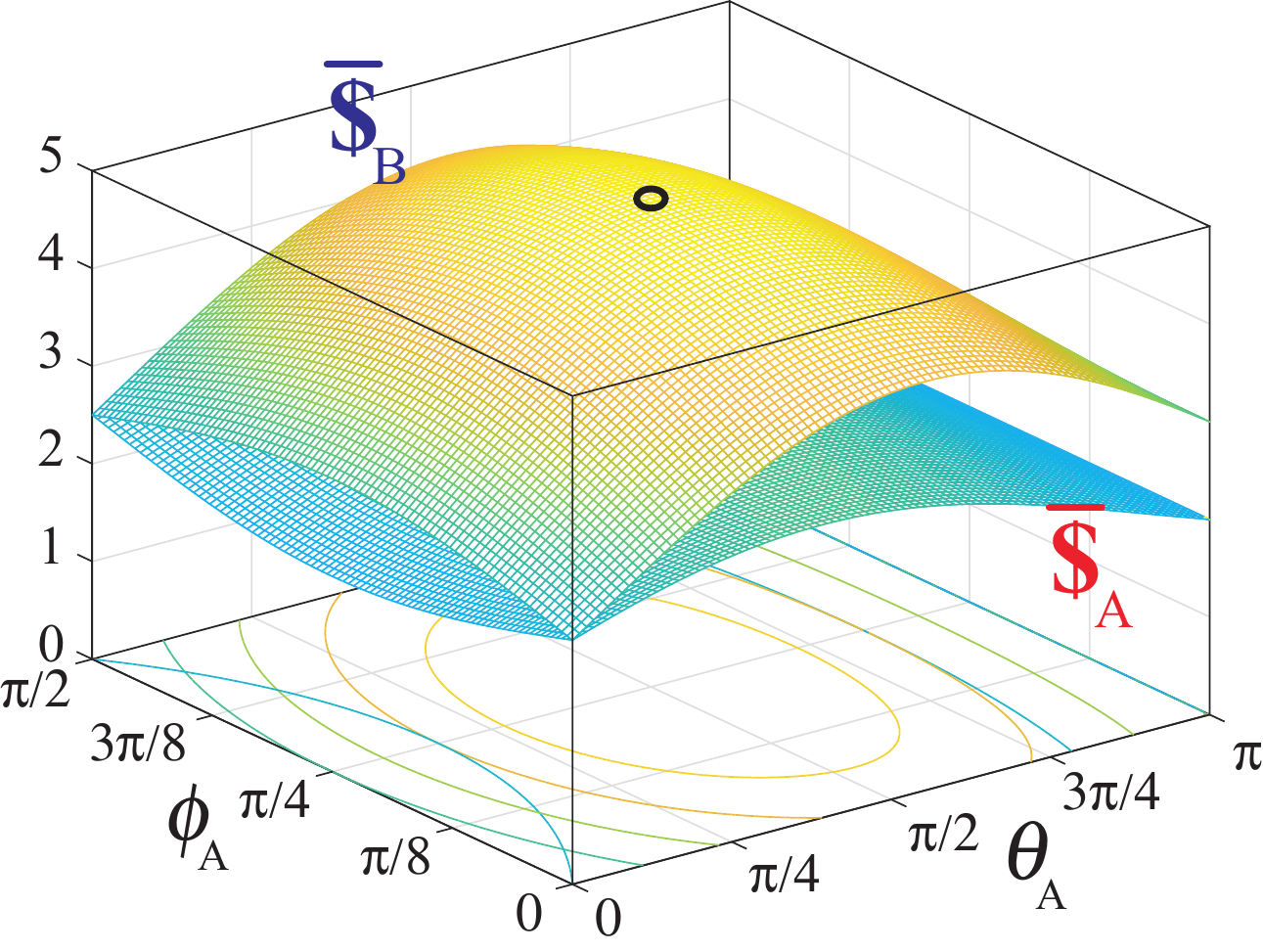}	\\ [0.00cm] 
\mbox{(d) $\phi_{\text{B}} =   0   \quad \theta_{\text{B}} = \pi/2 $ }   &   \mbox{(e) $\phi_{\text{B}} = \pi/8 \quad \theta_{\text{B}} = \pi/2 $ }   &   \mbox{(f) $\phi_{\text{B}} = \pi/4 \quad \theta_{\text{B}} = \pi/2 $ }	\\ [0.00cm] 
\epsfxsize=1.50in \epsffile{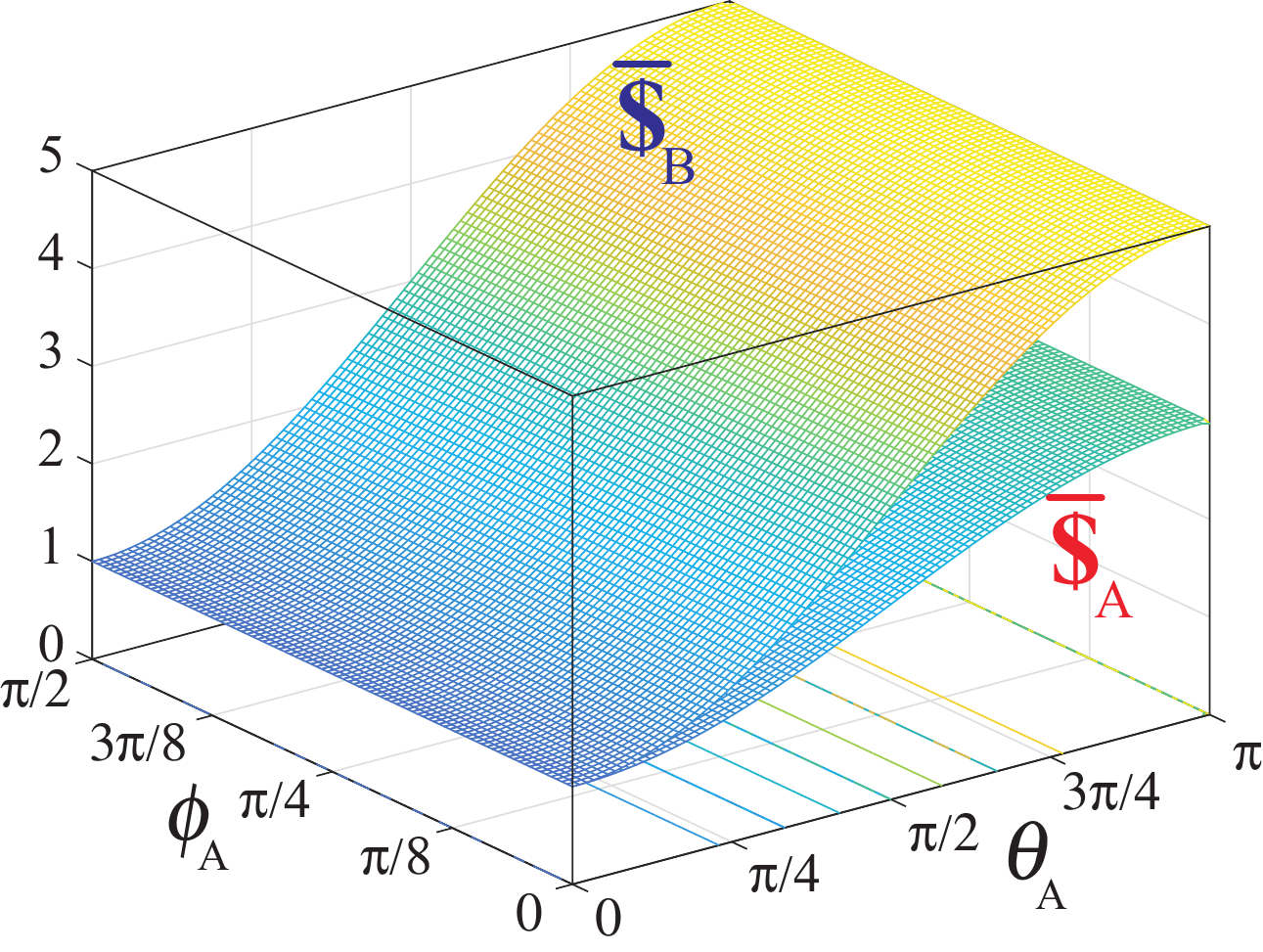}	& 
\epsfxsize=1.50in \epsffile{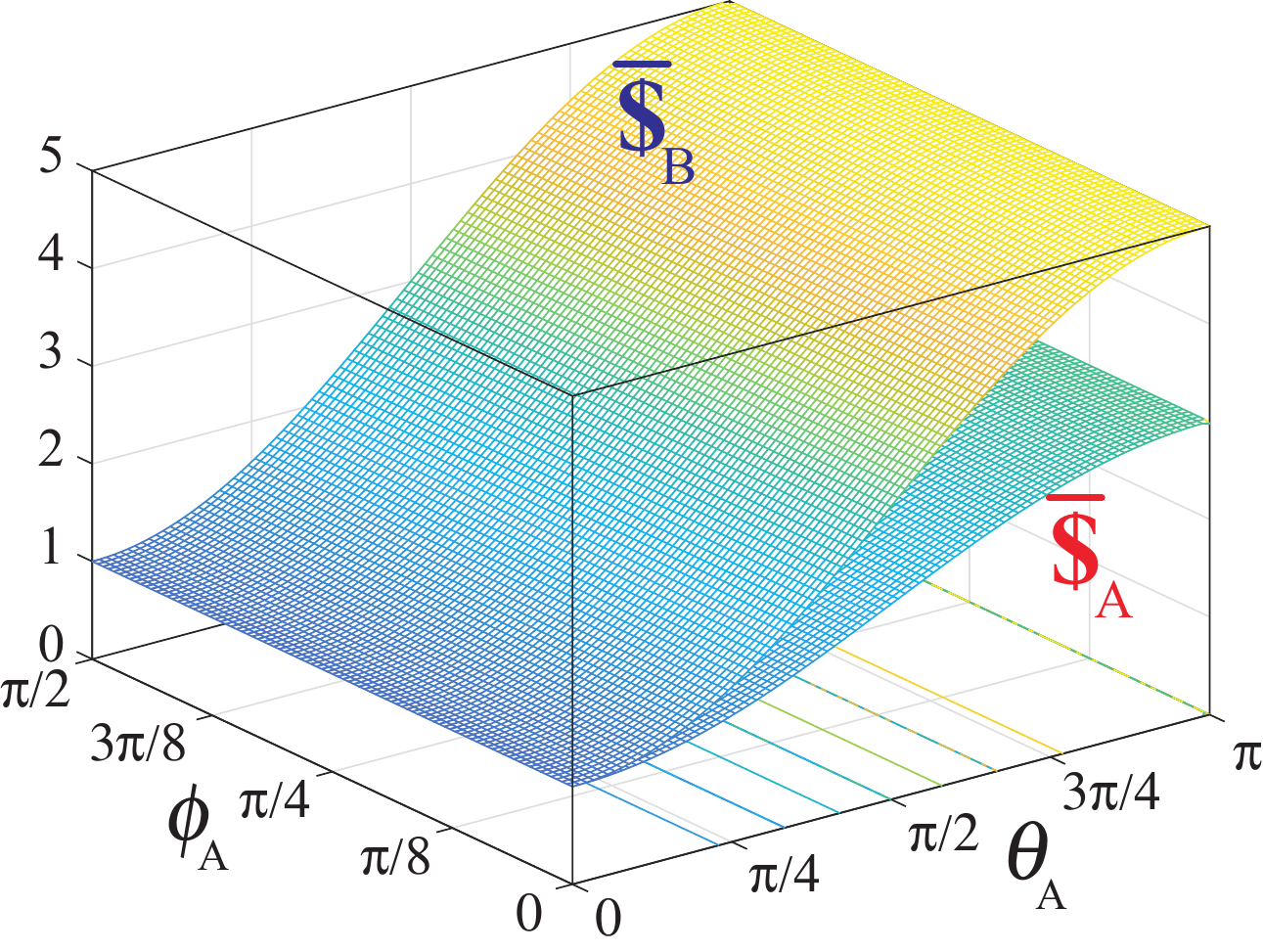}	& 
\epsfxsize=1.50in \epsffile{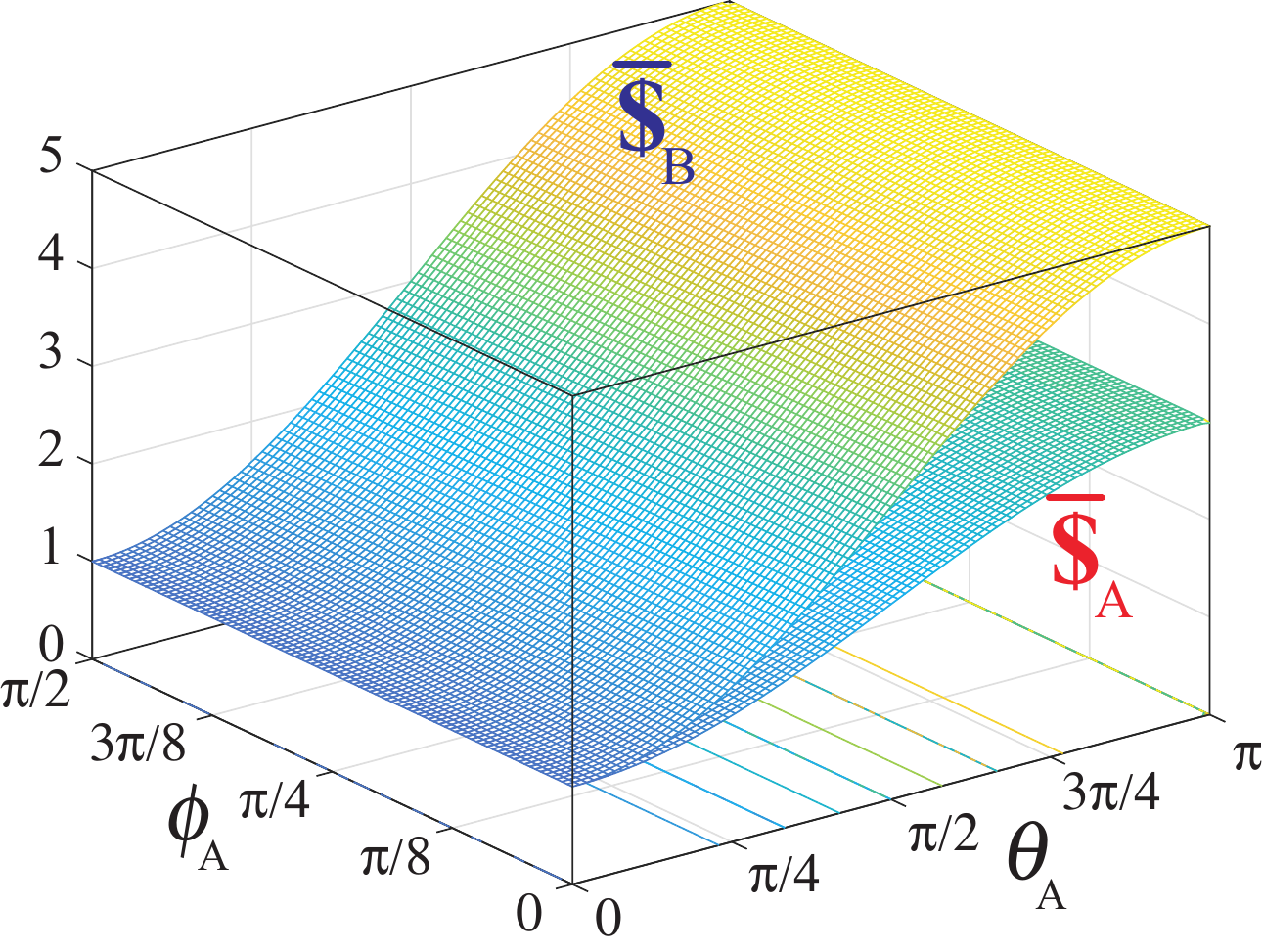}	\\ [0.00cm] 
\mbox{(g) $\phi_{\text{B}} =   0   \quad \theta_{\text{B}} = \pi $ }   &   \mbox{(h) $\phi_{\text{B}} = \pi/8 \quad \theta_{\text{B}} = \pi $ }   &   \mbox{(j) $\phi_{\text{B}} = \pi/4 \quad \theta_{\text{B}} = \pi $ }
\end{array}
$
\end{center}
\caption{ (Color online)  Surface plots of Alice's $\left( \overline{\$}_{\text{A}}  \right)$  and Bob's  $\left( \overline{\$}_{\text{B}}  \right)$  payoffs for the battle of the sexes quantum game,  {where $\alpha=5$,   $\beta=3$ and  $\gamma=1$ such that obey the relation $\alpha > \beta > \gamma $.} Surface plots are depicted under the  parameter domain  $\frac{\pi}{2} \geq  \phi_{\text{A}}  \geq 0$ and   $ \pi \geq  \theta_{\text{A}}  \geq  0$ such that the  angular parameters are setting to the following values $\phi_{\text{B}} =\left\lbrace 0 , \frac{\pi}{8}, \frac{\pi}{4}  \right\rbrace  $ and $\theta_{\text{B}} =\left\lbrace 0 , \frac{\pi}{2},\pi  \right\rbrace  $ .} \label{fig:ManyAngularParametersPayoffs}
\end{figure}

{It is interesting to visualize the information from the mathematical procedure detailed above  using 3D plots. In Fig.  \ref{fig:ManyAngularParametersPayoffs}, we display surface plots of the player's payoffs as a function of $\left( \phi_{ \text{A} }, \theta_{ \text{A} } \right)$ with the fixed values of  $ \phi_{ \text{B} } = 0, \frac{\pi}{8}, \frac{\pi}{4}  $ and  $\theta_{ \text{B} } = 0,  \frac{\pi}{2}, \pi $. Values of $\phi_{ \text{B} } =  \frac{3\pi}{8}   ,  \frac{\pi}{2}   $ would produce similar surfaces to  $\phi_{ \text{B} } =  \frac{\pi}{8} , 0 $, respectively, and are therefore not shown here. We would also produce similar surfaces if instead of fixing the values of   $ \left( \theta_{ \text{B} } , \phi_{ \text{B} } \right) $ and varying     $ \left( \theta_{ \text{A} } , \phi_{ \text{A} } \right) $, we fixed the values of    $ \left( \theta_{ \text{A} } , \phi_{ \text{A} } \right) $ and varied     $ \left( \theta_{ \text{B} } , \phi_{ \text{B} } \right) $.   Some of those parameters  are set to highlight the information generated from the mathematical procedure for $\phi_{ \text{A} } + \phi_{ \text{B} }  =  \frac{\pi}{4}$ related to the outcome of the mathematical procedure for the lower bound. In the particular case of $ \phi_{ \text{A} } = \phi_{ \text{B} } = \pi/8$ and $ \theta_{ \text{A} } = \theta_{ \text{B} } = 0$, it is represented by a black square symbol under the surface plot of Fig. \ref{fig:ManyAngularParametersPayoffs}(b). That point would be equivalent  to the point obtained with  $ \phi_{ \text{A} } + \phi_{ \text{B} } = \frac{3\pi}{4}$, with  $ \phi_{ \text{A} } = \phi_{ \text{B} } =  \frac{3\pi}{8}  $ and $ \theta_{ \text{A} } = \theta_{ \text{B} } = 0$. The result obtained at the upper bound is set by the linear equation  $ \phi_{ \text{A} } + \phi_{ \text{B} } =  \frac{\pi}{2} $ and   $ \tan \frac{\theta _{\text{A}}}{2}  =  \cot \frac{\theta _{\text{B}}}{2}$. From these  equations, the trivial angular parameters are given by $\phi_{ \text{A} } = \phi_{ \text{B} } =   \frac{\pi}{4}  $ and $ \theta _{\text{A}} = \theta _{\text{B}} =  \frac{\pi}{2} $, and represented by a black circle symbol under the surface plot of  Fig. \ref{fig:ManyAngularParametersPayoffs}(f). }

{Therefore, the main information generated by the choice of the angular parameters following the mathematical analysis introduced in this appendix are: (I) Increasing (Decreasing) the  values of  $ \theta _{ \text{A,B} } $ inhibits (promotes) the effects on the values of  $ \phi _{ \text{A,B} } $; (II) The lower bounds generated from the analysis of Eq. (\ref{BhaskarasSolutionCorrelated}) allows identifying angular parameter values that are checked by two linear equations, where the set of trivial values are  $\left\{ \phi _{\text{A}}=\phi _{\text{B}}=\frac{\pi }{8},\theta _{\text{A} } = \theta _{\text{B}}=0\right\} $ and $\left\{ \phi _{\text{A}}=\phi _{\text{B}}=\frac{3\pi }{8},\theta _{\text{A} } = \theta _{\text{B}}=0\right\} $. Similarly, the upper bound allows identifying the other set of trivial values $ \left\{ \phi _{\text{A}}=\phi _{\text{B}}=\frac{\pi }{4},\theta _{\text{A}   } = \theta _{\text{B}}=\frac{\pi }{2}\right\} $; (III) The bounds related to  Eq. (\ref{BhaskarasSolutionMas}) and Eq. (\ref{BhaskarasSolutionMenos}) have trivial angular values given by   $ \left\{ \phi _{\text{A}}=\phi _{\text{B}} = 0 ,\theta _{\text{A}   } = \theta _{\text{B}}=\frac{\pi }{2}\right\} $ and $ \left\{ \phi _{\text{A}}=\phi _{\text{B}}=\frac{\pi }{2},\theta _{\text{A}   } = \theta _{\text{B}}=\frac{\pi }{2}\right\} $. (IV) The information about the parameter values generated from Eq. (\ref{BhaskarasSolutionCorrelated}) for positive (negative) real values of $\Phi$ corresponds to high (low) correlation between payoffs values of each player.}

With all these equations, we can now analyse the payoff functions  to quantify the degree of  happiness  of each player under the following:
\begin{enumerate}
\item[1$^{\circ}$] {The first set of parameter values   $ \theta_{\text{A}} = \theta_{\text{B}} = 0$ and  $ \phi_{\text{A}} + \phi_{\text{B}} = \frac{\pi}{4} $    define the quantum strategy  operator $\widehat{\mathbf{U}}\left( 0, \phi_{\text{A}}  \right) \otimes \widehat{\mathbf{U}}\left(0, \phi_{\text{B}}  \right) $. The payoffs of both players are $ \overline{\$}_{\text{A}} = \frac{\alpha+ \beta}{2}$ and $ \overline{\$}_{\text{B}} = \frac{\beta + \alpha}{2} $ which when substituting the values for $\alpha$, $\beta$ and $\gamma$ the payoffs becomes  $ \overline{\$}_{\text{A}} \left(  \theta _{\text{A}}  = 0 ,\theta _{\text{B}} = 0,\phi_{\text{A}}, \phi_{\text{B}} \right) =  \overline{\$}_{\text{B}} \left(  \theta _{\text{A}}  = 0 ,\theta _{\text{B}} = 0,\phi_{\text{A}}, \phi_{\text{B}} \right) = 4 $.}
\item[2$^{\circ}$] {The second set of parameter values  $ \theta_{\text{A}} = \theta_{\text{B}} = 0$ and  $ \phi_{\text{A}} + \phi_{\text{B}} = \frac{3\pi }{4} $, which defines  $\widehat{\mathbf{U}}\left( 0,  \phi_{\text{A}}  \right) \otimes \widehat{\mathbf{U}}\left( 0,  \phi_{\text{B}} \right) $, so that the payoffs of both players are $ \overline{\$}_{\text{A}} = \frac{\alpha+ \beta}{2}$ and $ \overline{\$}_{\text{B}} = \frac{\beta + \alpha}{2} $. Substituting the values for $\alpha$ and $\beta$ the payoffs assume the value  $ \overline{\$}_{\text{A}} \left(  \theta _{\text{A}}  = 0 ,\theta _{\text{B}} = 0,\phi_{\text{A}}, \phi_{\text{B}} \right) =  \overline{\$}_{\text{B}} \left(  \theta _{\text{A}}  = 0 ,\theta _{\text{B}} = 0,\phi_{\text{A}}, \phi_{\text{B}} \right) = 4 $.}
\item[3$^{\circ}$] {The third  set of parameter values is    $ \tan \frac{\theta _{\text{A}}}{2}  =  \cot \frac{\theta _{\text{B}}}{2}$  and $ \phi_{\text{A}} + \phi_{\text{B}} =  \frac{\pi }{2} $, which establish the quantum strategy operator    $\widehat{\mathbf{U}}\left( \theta _{\text{A}},  \phi_{\text{A}} \right) \otimes \widehat{\mathbf{U}}\left( \theta _{\text{B}} ,  \phi_{\text{B}} \right) $. The payoffs of both players are $ \overline{\$}_{\text{A}} = \gamma \left( \left( \cos \frac{\theta _{\text{A}%
}}{2}\sin \frac{\theta _{\text{B}}}{2}\right) ^{2}+\left( \sin \frac{\theta
_{\text{A}}}{2}\cos \frac{\theta _{\text{B}}}{2}\right) ^{2}-\frac{\sin
\theta _{\text{A}}\sin \theta _{\text{B}}}{2}\right)  +\beta \left( \sin \frac{\theta _{\text{A}}}{2}\sin \frac{\theta _{\text{B}%
}}{2}+\cos \frac{\theta _{\text{A}}}{2}\cos \frac{\theta _{\text{B}}}{2}%
\right) ^{2}$ and $ \overline{\$}_{\text{B}} = \gamma \left( \left( \cos \frac{\theta _{\text{A}%
}}{2}\sin \frac{\theta _{\text{B}}}{2}\right) ^{2}+\left( \sin \frac{\theta
_{\text{A}}}{2}\cos \frac{\theta _{\text{B}}}{2}\right) ^{2}-\frac{\sin
\theta _{\text{A}}\sin \theta _{\text{B}}}{2}\right)  +\alpha \left( \sin \frac{\theta _{\text{A}}}{2}\sin \frac{\theta _{\text{B%
}}}{2}+\cos \frac{\theta _{\text{A}}}{2}\cos \frac{\theta _{\text{B}}}{2}%
\right) ^{2}$, which when substituting the numerical values for  $\alpha$, $\beta$ and $\gamma$ at their maximum angular parameter values yields   $ \overline{\$}_{\text{A}}\left(  \theta _{\text{A}}=\frac{\pi }{2},\theta _{\text{B}}=\frac{\pi }{2}, \phi_{\text{A}}, \phi_{\text{B}} \right)  = 3 $ and $  \overline{\$}_{\text{B}}\left(  \theta _{\text{A}}=\frac{\pi }{2},\theta _{\text{B}}=\frac{\pi }{2},\phi_{\text{A}}, \phi_{\text{B}} \right) = 5 $.}
\item[4$^{\circ}$] {The fourth  set of parameter values is   $ \tan \frac{\theta _{\text{A}}}{2}  =  \cot \frac{\theta _{\text{B}}}{2}$    and $ \phi_{\text{A}} + \phi_{\text{B}} = 0 $, which establish the quantum strategy operator    $\widehat{\mathbf{U}}\left( \theta _{\text{A}}  , \phi_{\text{A}} \right) \otimes \widehat{\mathbf{U}}\left( \theta _{\text{B}} , \phi_{\text{B}} \right) $. The payoffs of both players are $ \overline{\$}_{\text{A}} = \frac{\alpha+2 \gamma + \beta}{4}$ and $ \overline{\$}_{\text{B}} = \frac{\beta + 2 \gamma + \alpha}{4} $, which when substituting the numerical values for  $\alpha$, $\beta$ and $\gamma$ yields   $ \overline{\$}_{\text{A}}  \left(  \theta _{\text{A}} ,\theta _{\text{B}},\phi_{\text{A}}, \phi_{\text{B}} \right) =  \overline{\$}_{\text{B}} \left(  \theta _{\text{A}} ,\theta _{\text{B}},\phi_{\text{A}}, \phi_{\text{B}} \right) = 2.5 $.}
\item[5$^{\circ}$] {The fifth    set of parameter values is    $ \tan \frac{\theta _{\text{A}}}{2}  =  \cot \frac{\theta _{\text{B}}}{2}$    and $ \phi_{\text{A}} + \phi_{\text{B}} = \pi $, which establish the quantum strategy operator   $\widehat{\mathbf{U}}\left( \theta _{\text{A}}  , \phi_{\text{A}} \right) \otimes \widehat{\mathbf{U}}\left( \theta _{\text{B}} , \phi_{\text{B}} \right) $. The payoffs of both players are $ \overline{\$}_{\text{A}} = \frac{\alpha+2 \gamma + \beta}{4}$ and $ \overline{\$}_{\text{B}} = \frac{\beta + 2 \gamma + \alpha}{4} $, which when substituting the numerical values for  $\alpha$, $\beta$ and $\gamma$ yields   $ \overline{\$}_{\text{A}}  \left(  \theta _{\text{A}} ,\theta _{\text{B}},\phi_{\text{A}}, \phi_{\text{B}} \right) =  \overline{\$}_{\text{B}} \left(  \theta _{\text{A}} ,\theta _{\text{B}},\phi_{\text{A}}, \phi_{\text{B}} \right) = 2.5 $.}
\end{enumerate}

The set of angular parameters improves the degree of happiness of both players by taking advantage of the quantum mechanical nature of the game \cite{lee2003}, and these strategies have no classical counterpart \cite{van-Enk2002}. To make it more evident, at the left and right column of Fig.  \ref{fig:DensityMatrices}  we plot the theoretical density matrix elements as bar charts. The second, fourth and sixth bar charts represent the quantum state generated by performing quantum operator strategies using the above angular parameters. The fourth and sixth bar charts represent quantum states with maximum quantum correlations {studied using Quantum discord \cite{auccaise2011A,auccaise2011B} or entropic methodologies \cite{deng2018}}. {In this sense, the fourth density matrix characterised by the parameters $\theta_{\text{A,B}}=0$ and $\phi_{\text{A.B}}=\frac{\pi}{8}$ is equivalent to the Bell state $\left\vert \Phi^{+}_{\text{ap}} \right\rangle$ as sketched on Fig. 2 of Ref. \cite{reiserer2014}. Also, the sixth density matrix with parameters   $\theta_{\text{A,B}}=0$ and $\phi_{\text{A,B}}=\frac{3\pi}{8}$  is equivalent to the Bell state  $\left\vert \Phi^{-}_{\text{ap}} \right\rangle$. } The first, third, fifth and seventh bar charts represent quantum states without quantum correlations, which could, therefore, have classical counterpart.

\end{appendices}

\end{document}